\documentclass{JHEP3}    
\usepackage{epsfig}

\newcommand{\fref}[1]{Figure~\ref{#1}}
\newcommand{\cref}[1]{Chapter~\ref{#1}}
\newcommand{\beq}{\begin{equation}}
\newcommand{\eeq}{\end{equation}}
\newcommand{\ba}{\begin{array}}
\newcommand{\ea}{\end{array}}
\newcommand{\bcenter}{\begin{center}}
\newcommand{\ecenter}{\end{center}}

\def\IB{\relax\hbox{$\inbar\kern-.3em{\rm B}$}}
\def\IC{\relax\hbox{$\inbar\kern-.3em{\rm C}$}}
\def\ID{\relax\hbox{$\inbar\kern-.3em{\rm D}$}}
\def\IE{\relax\hbox{$\inbar\kern-.3em{\rm E}$}}
\def\IF{\relax\hbox{$\inbar\kern-.3em{\rm F}$}}
\def\IG{\relax\hbox{$\inbar\kern-.3em{\rm G}$}}
\def\IGa{\relax\hbox{${\rm I}\kern-.18em\Gamma$}}
\def\IH{\relax{\rm I\kern-.18em H}}
\def\IK{\relax{\rm I\kern-.18em K}}
\def\IL{\relax{\rm I\kern-.18em L}}
\def\IP{\relax{\rm I\kern-.18em P}}
\def\IR{\relax{\rm I\kern-.18em R}}
\def\IZ{\relax\ifmmode\mathchoice
{\hbox{\cmss Z\kern-.4em Z}}{\hbox{\cmss Z\kern-.4em Z}}
{\lower.9pt\hbox{\cmsss Z\kern-.4em Z}}
{\lower1.2pt\hbox{\cmsss Z\kern-.4em Z}}\else{\cmss Z\kern-.4em Z}\fi}
\def\II{\relax{\rm I\kern-.18em I}}

\def\sCC{{\kern 0.27em\vrule height1.45ex width0.03em depth0em
          \kern-0.30em\rm C}}
\def\C{{\mathchoice
  {\sCC}
  {\sCC}
  {\kern 0.225em \vrule height1.05ex width0.025em depth0em \kern-0.25em \rm C}
  {\kern 0.180em \vrule height0.78ex width0.02em depth0em \kern-0.2em \rm C}
        }}
\def\sHH{{\rm I\kern-.16em{}H}}
\def\H{{\mathchoice
  {\sHH}
  {\sHH}
  {\rm I\kern-.13em{}H}
  {\rm I\kern-.13em{}H} }}
\def\sNN{{\rm I\kern-.16em{}N}}
\def\N{{\mathchoice
  {\sNN}
  {\sNN}
  {\rm I\kern-.12em{}N}
  {\rm I\kern-.10em{}N} }}
\def\sPP{{\rm I\kern-.16em{}P}}
\def\P{{\mathchoice
  {\sPP}
  {\sPP}
  {\rm I\kern-.12em{}P}
  {\rm I\kern-.10em{}P} }}
\def\sQQ{{\kern 0.27em \vrule height1.45ex width0.03em depth0em
          \kern-0.30em \rm Q}}
\def\Q{{\mathchoice
        {\sQQ}
        {\sQQ}
  {\kern 0.225em \vrule height1.05ex width0.025em depth0em \kern-0.25em \rm Q}
  {\kern 0.180em \vrule height0.78ex width0.020em depth0em \kern-0.20em \rm Q}
        }}
\def\sRR{{\rm I\kern-0.16em{}R}}
\def\R{{\mathchoice
  {\sRR}
  {\sRR}
  {\rm I\kern-0.12em{}R}
  {\rm I\kern-0.10em{}R} }}
\def\sZZ{{\rm Z\kern-0.32em{}Z}}
\def\Z{{\mathchoice
  {\sZZ}
  {\sZZ} 
  {\rm Z\kern-0.3em{}Z}     
  {\rm Z\kern-0.25em{}Z} }}  
\def\ZZZ{{\rm Z\kern-0.24em{}Z}}
\def\sII{{\rm I\kern-0.16em{}I}}
\def\I{{\mathchoice
  {\sII}
  {\sII}
  {\rm I\kern-0.12em{}I}
  {\rm I\kern-0.10em{}I} }}

\def\inbar{\,\vrule height1.5ex width.4pt depth0pt}
\font\cmss=cmss10 \font\cmsss=cmss10 at 7pt

\def\smiley{\hbox{\large$\bigcirc$\hspace{-0.80em}\raise.2ex
\hbox{$\cdot\cdot$}\kern-.61em\lower.2ex\hbox{\scriptsize$\smile$}}\ }
\def\frowny{\hbox{\large$\bigcirc$\hspace{-0.80em}\raise.2ex
\hbox{$\cdot\cdot$}\kern-.635em\lower.2ex\hbox{\scriptsize$\frown$}}\ }

\def\I{{\rlap{1} \hskip 1.6pt \hbox{1}}}

\makeatletter
\let\hangafter\@hangfrom
\makeatother

\newcommand{\be}{\begin{equation}}
\newcommand{\ee}{\end{equation}}
\newcommand{\bea}{\begin{eqnarray}}
\newcommand{\eea}{\end{eqnarray}}
\newcommand{\bean}{\begin{eqnarray*}}
\newcommand{\eean}{\end{eqnarray*}}

\preprint{MIT-CTP-3286\\ \\ {\tt hep-th/0207006}}

\title{Geometric dualities in 4d field theories and their 5d interpretation}

\author{
Sebasti\'{a}n Franco and Amihay Hanany
\footnote{
Research supported in part by the CTP and the LNS
of MIT and the U.S. Department of Energy under cooperative agreement
$\#$DE-FC02-94ER40818. A. H. is also supported by the Reed Fund Award and 
a DOE OJI award.}
\\
~\\
Center for Theoretical Physics,
\\ Massachusetts Institute of Technology,\\
Cambridge, MA 02139, USA.\\
\email{sfranco, hanany@mit.edu}
}

\abstract{We study four-dimensional ${\cal N}=1$ gauge theories arising on D3-branes probing toric singularities. Toric dualities
and flows between theories corresponding to different singularities are analyzed by encoding the geometric information into
$(p,q)$ webs. A new method for identifying quiver symmetries of the four-dimensional theories 
using the brane webs is developed. Five-dimensional theories are associated
to the theories on the D3-branes by using $(p,q)$ webs. This leads to a novel interpretation of Seiberg duality, which can be mapped to the crossing 
of curves of marginal stability in five dimensions.}
\begin{document}

\section{Introduction}

String theory has been widely used to study the dynamics of gauge theories. In doing so, it sometimes provides a completely 
new interpretation of field theory results. The relation goes in both directions, and gauge theories can be used to understand 
string theory processes. The main ingredient in this connection is the fact that the low energy 
dynamics of D-branes is described by SYM on their world volume, with different amounts of supersymmetry depending on the specific configuration. 
Several ways of constructing gauge theories using D-branes have been developed. The main strategies employed are brane setups \cite{HW}, 
geometric engineering \cite{geomeng} and brane probes \cite{braneprobes}. 

Toric duality was discovered while studying the gauge theories arising on D-branes probing toric singularities. The classical moduli 
space of a D-brane on a singularity is exactly the probed geometry \cite{Witten}. In \cite{toric,phases}, it was discovered 
that given a toric singularity, the corresponding gauge theory is not unique. The different phases associated to a singularity 
form equivalence classes of ${\cal N}=1$  four dimensional models that describe the same physics in the IR limit. In \cite{dual,Chris2,Vafa}, it was 
shown that toric duality is Seiberg duality. In this way, highly non-trivial sets of dual theories were constructed, 
and Seiberg duality acquired a fully geometric interpretation in string theory.

\begin{figure}[h]
  \epsfxsize = 10cm
  \centerline{\epsfbox{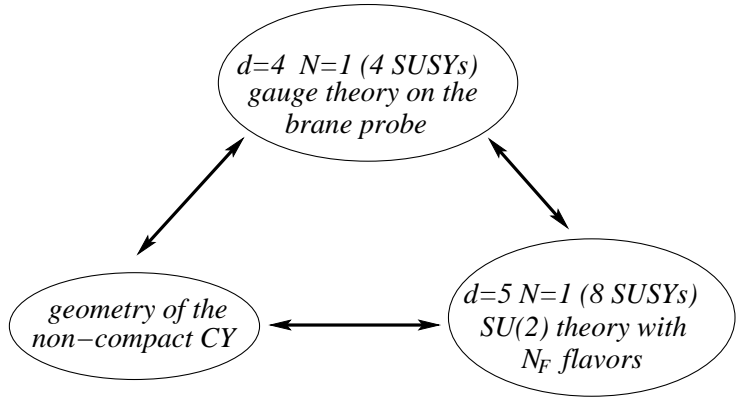}}
  \caption{The three alternative perspectives that will be developed in this paper. The connections between them will be made using $(p,q)$ webs.}
  \label{viewpoints}
\end{figure}

The purpose of this paper is to look at gauge theories living on D3-branes probing toric singularities from three completely different 
perspectives, studying not only the transition between toric dual theories but also the flow between theories corresponding to different 
geometries. These complementary approaches are summarized in \fref{viewpoints}, and we will discuss how different physical processes manifest in the 
three languages. In the first place, we will study directly the ${\cal N}=1$ theories 
in $d=4$. In this context, toric duals are related by Seiberg dualities, while theories for different geometries correspond to (un)higgsings. 
The second viewpoint is purely geometric, and the continuous flow between theories is achieved by blow-ups 
and blow-downs of the corresponding non-compact Calabi-Yau. Finally, every theory under study has an associated five dimensional 
${\cal N}=1$, $SU(2)$ partner \footnote{Five dimensional ${\cal N}=1$ theories have 8 supercharges. This is 
the number of SUSYs that is preserved by the $(p,q)$ web configuration when condition 
\ref{slope} is satisfied.}. The correspondence follows from considering $M$ theory on the different CY threefolds. In this language, there exist
a one to one mapping between the change of parameters that interpolates between four dimensional toric dual theories 
and the change of the BPS spectrum in five dimensions (crossing of curves of marginal stability). The key objects interconnecting these three descriptions are $(p,q)$ webs.

The organization of this paper is as follows. In Section 2, we review the main concepts of  $(p,q)$ web constructions and toric geometry. In Section 3, 
we explain how to extract 
the quiver for the four dimensional theory that appears in the world volume of D3-branes probing a toric variety whose toric data is 
encoded in a given $(p,q)$ web. Section 4 is devoted to understanding the flow between toric duals and theories associated to different
singularities as geometric transitions and (un)higgsings. Section 5 shows the full power 
of $(p,q)$ webs in establishing quiver symmetries
of the four dimensional gauge theories. In section 6, we use the mapping from $(p,q)$ webs
to five dimensional theories to introduce a third perspective for toric duality and geometric
transitions, namely there is a one to one correspondence between the continuous change of parameters and subsequent flops in $(p,q)$ webs 
which is associated to toric duality and a change in the five dimensional BPS spectrum.   

\section{A review of $(p,q)$ webs and toric geometry}

\subsection{$(p,q)$ webs and five dimensional theories}

In \cite{webs1,webs2} $(p,q)$ webs were introduced as brane constructions to study five dimensional theories. They are Type IIB 
string theory configurations in which 5-branes share four spatial directions and time. The remaining dimension of the 
5-branes world volumes lie on a plane parametrized by the $(x,y)$ coordinates. Every $(p,q)$ web can be associated to a ${\cal N}=1$ 
gauge theory living in the $4+1$ dimensions common to all the 5-branes. Each brane has a $(p,q)$ charge which 
is related to its tension

\beq
T_{p,q}=|p+\tau q|T_{D_5}
\label{tension}
\eeq
and to its slope

\beq
\Delta x: \Delta y=p:q
\label{slope}
\eeq
where $T_{D_5}$ is the D5-brane tension and $\tau$ is the complex scalar of Type IIB (which 
we have chosen equal to $i$ in \ref{slope}). The last condition assures that 1/4 of
the SUSYs is preserved. Branes can join at vertices, where $(p,q)$ charge is conserved,

\beq
\sum_i p_i=\sum_i q_i=0
\label{conservation}
\eeq
where the sum is performed over all the branes ending at a given vertex. It is easy to see that \ref{tension}, \ref{slope} and 
\ref{conservation} imply the equilibrium of the web.

\begin{figure}[h]
  \epsfxsize = 10cm
  \centerline{\epsfbox{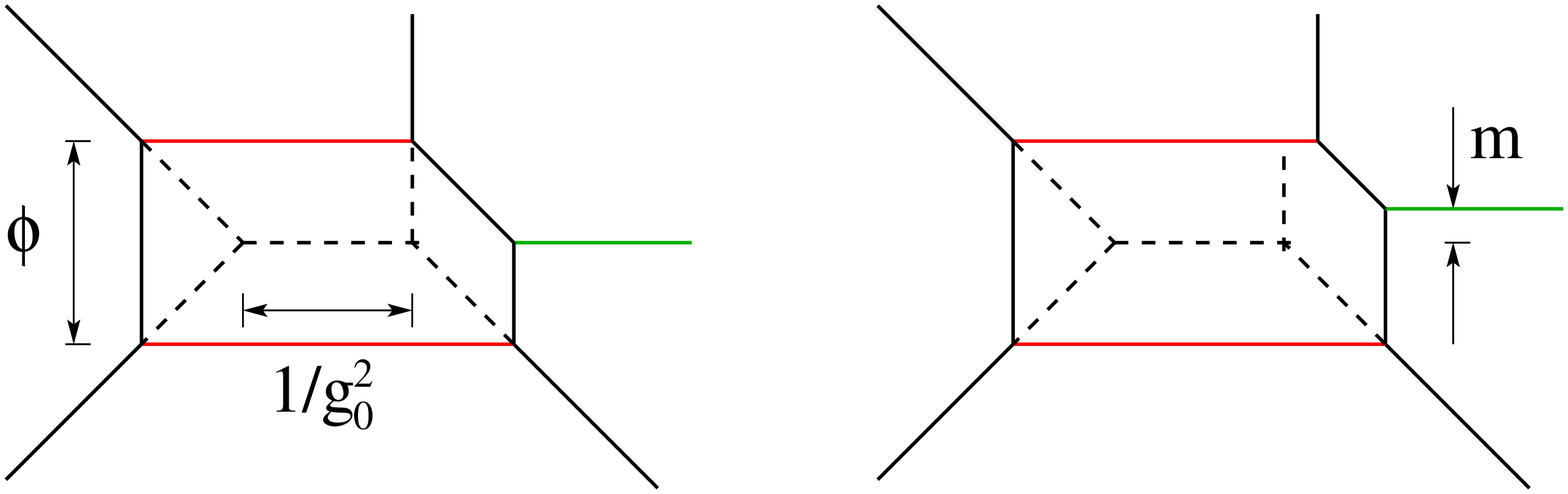}}
  \caption{A $(p,q)$ web corresponding to an $SU(2)$ theory with one flavor.}
  \label{5_dimensional}
\end{figure}

These theories were thoroughly studied in \cite{webs2}. We will give here a brief explanation on how the five dimensional parameters
can be read off from the $(p,q)$ webs. All along this paper we will deal with $SU(2)$ theories, so we choose an $SU(2)$ 
model with one flavor to exemplify all the relevant concepts (\fref{5_dimensional}). Color branes are finite parallel 
branes (depicted in red in the figure). Their separation is parametrized by the expectation value of a $U(1)$ scalar 
$\phi$. For $\phi=0$, both color branes coincide and we have an unbroken $SU(2)$ gauge symmetry.
When $\phi>0$, $SU(2)$ is broken down to $U(1)$ and the $W$ gauge boson gets a mass $m_W=\phi$. The bare value of the gauge coupling is given by
the length of the color branes when $\phi=0$ (see \fref{5_dimensional})

We can add a flavor to this theory. This is represented by a semi-infinite brane parallel to the color branes (green brane in \fref{5_dimensional}).
For $\phi=0$ this brane corresponds to a quark multiplet in the {\bf 2} representation of $SU(2)$. When $\phi$ grows, the gauge group is higgsed
to $U(1)$ and the {\bf 2} gives rise to two quark states with $m_{1,2}=\phi/2$. A bare mass for the quarks can
be introduced by displacing the flavor brane with respect to the middle position between the color branes. In this case, the quark masses become 
$m_{1,2}=|m\pm \phi/2|$.

BPS saturated states correspond to string webs ending perpendicularly on the five branes. The rules governing the construction of string webs are identical to the ones
we discussed for brane webs. The monopole tension is calculated as the 
area of the closed face of the web. The masses of BPS states and the monopole tension can be expressed in terms of 
$\phi$, $g_0$ and the quark bare masses.

\bigskip

\subsection{Toric geometry}

Toric geometry and $(p,q)$ webs are closely related. Toric geometry studies varieties that admit a $U(1)^d$ action, 
in general with fixed points (for a complete treatment look at \cite{Fulton}). These spaces are described by specifying shrinking cycles and relations between them. 
An alternative description of this geometries is in terms of $(p,q)$ webs. It is possible to see that the connection 
between both descriptions consists simply in that the brane web is a representation of the toric skeleton (for
a complete discussion of the relation see \cite{leung_vafa}). 

In this paper we will focus on cones over two complex dimensional toric varieties. They can be 
understood as $T^2$ fibrations over $\IC$. Lines and vertices in the web represent fixed points of the $U(1)$ 
actions (i.e. places where the fibration becomes degenerate). Lines correspond to vanishing 1-cycles, 
while points describe vanishing 2-cycles. The process of blowing-up a point in a given variety consists of replacing 
it by a 2-sphere. The toric representation of a 2-sphere is a segment. Then we see that if the points we choose to 
blow up are vertices in the web, we just have to replace them by segments. 

\begin{figure}[h]
  \epsfxsize = 12cm
  \centerline{\epsfbox{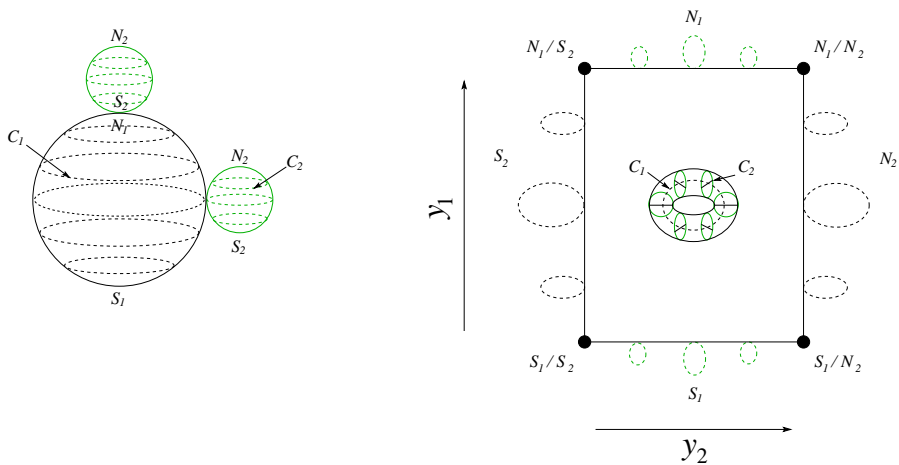}}
  \caption{Toric representation of $F_0=\IP^1 \times \IP^1$.}
  \label{toric}
\end{figure}

We conclude this brief introduction by developing the toric representation for a specific case, the zeroth Hirzebruch
surface $F_0$. $F_0$ is equal to $\IP^1 \times \IP^1$, so we can think about it as a 2-sphere fibered over another
2-sphere. This representation is shown in \fref{toric}.a. Now we want to interpret this geometry
as a $T^2$ fibration over $\IC$. A natural way to do this is by associating the vertical positions $y_{1,2}$ on the two 2-spheres to the 
two coordinates in the complex plane. Then, we associate to every point on $\IC$ a 2-torus given by 
the product of the two circles parallel to the equators at the corresponding $y_i$. The full construction is 
presented in \fref{toric}.b. The $C_2$ circle vanishes at the north and south poles of the small sphere, represented
torically by the two vertical lines. Analogously, the two horizontal lines correspond to the north and south poles of
the big sphere. Both $C_1$ and $C_2$ vanish at the four vertices of the rectangle. From this discussion we also see that
the sizes of the different compact 2-cycles are given by the lengths of the segments in the toric skeleton.

\section{Four dimensional quivers from $(p,q)$ webs}

Let us study how to obtain the quiver for the four dimensional ${\cal N}=1$ theory that appears on the world volume of a
D3-brane probing the type of singularities we are considering. A possible approach consists of obtaining the 
singularity as a partial resolution of an abelian orbifold, whose associated gauge theory is well understood. This approach
was pursued in \cite{toric} and further developed in \cite{phases,Chris2,symmetries} to get the theories for the zeroth Hirzebruch and the
toric del Pezzo surfaces as partial resolutions of $\IC^3/{(\IZ_3 \times \IZ_3)}$.

\begin{figure}[h]
\begin{center}
$
\begin{array}{c}
\begin{array}{ccccc}  \mbox{D3} & \ \ \ \ \ \ \ \ \ \ \ \ \ \ \ \ \ \ \ \ \ \ \ \ \ \ \ & \mbox{D5} & \ \ \ \ \ \ \ \ \ \ \ \ \ \ \ \ \ \ \ \ \ \ \ \ \ \ \ & \mbox{D7}\end{array} \\
\begin{array}{ccc} {\epsfxsize=4cm\epsfbox{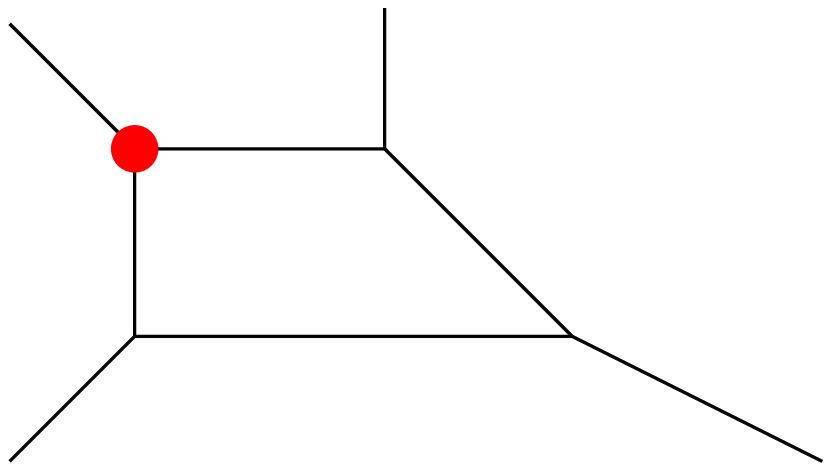}} &
		   {\epsfxsize=4cm\epsfbox{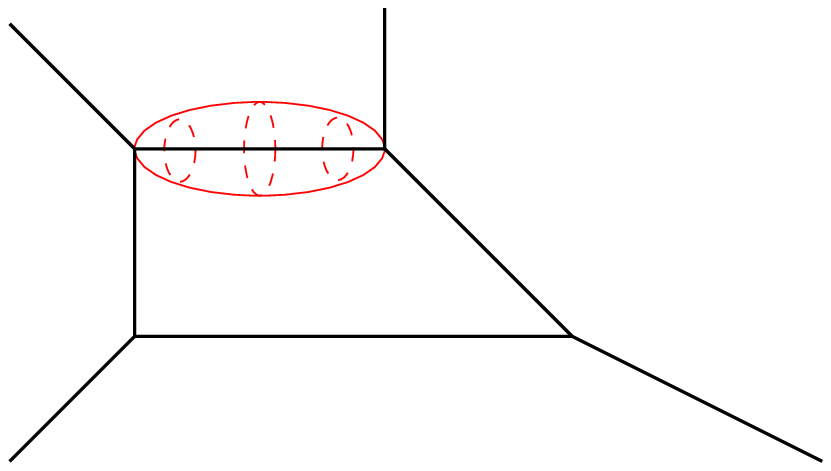}} &
                   {\epsfxsize=4cm\epsfbox{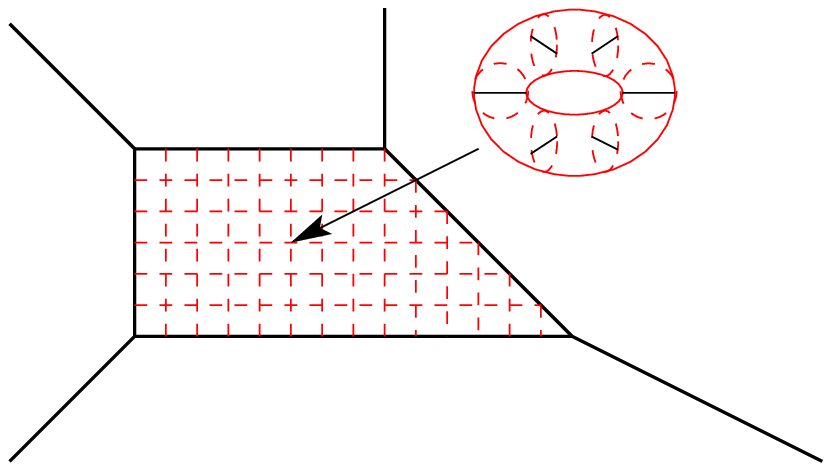}}  
\end{array}
\end{array}
$
\end{center}
\caption{Possible D3, D5 and D7-branes located at 0-cycles and wrapping compact 2 and 4 cycles, respectively.}
\label{fractional_1}
\end{figure}

A second alternative exploits the geometric information encoded in the $(p,q)$ web. Each factor of the gauge group is given by
a fractional brane, which is a bound state of D3, D5 and D7-branes. D3-branes span the four directions transverse to the singularity and
thus are located at 0-cycles inside the toric variety. Analogously, D5-branes wrap 2-cycles 
and D7-branes wrap the compact 4-cycle. Some possible configurations are sketched in 
\fref{fractional_1}. The mirror Type IIA geometries associated to these models were 
studied in \cite{HI}. It was found there that 0, 2 and 4-cycles map to
3-cycles, and D3-branes become D6-branes wrapping a ${\cal T}^3$. The bifundamental matter content is given by the intersection
matrix of the 3-cycles. Furthermore, each 3-cycle $S_i$ wraps a 1-cycle $C_i$ of a smooth 
elliptic fiber that becomes degenerate at some point $z_i$. Each $C_i$ carries a
$(p_i,q_i)$ charge, and the intersection numbers for the 3-cycles can be calculated as

\beq
^\# (S_i.S_j)= {}^\#(C_i.C_j)=\det \left( \begin{array}{cc} p_i & q_i \\ p_j & q_j \end{array} \right)
\label{intersection}
\eeq

The $(p,q)$ charges of the 1-cycles are those of the external legs of the web. This suggests a profound connection between the 
$(p,q)$ web and the gauge theory in four dimensions. Each node in the web corresponds to the fractional brane of one gauge group
factor. The external leg associated to it gives the $(p,q)$ charges of the 1-cycle in the mirror manifold used to compute the
matter content using the intersections with other 1-cycles. 

As it has already been noticed in \cite{HI}, $(p,q)$ charge conservation at every node of the web ( $\sum_i (p_i,q_i)=0$) guarantees the absence of anomalies in the four dimensional gauge theory. 
This is the case if every node of the quiver has same number of incoming and outcoming arrows. Choosing the i-th node, we have

\begin{eqnarray}
N^{(i)}_{in}-N^{(i)}_{out}=\sum_{j\neq i} \det \left( \begin{array}{cc} p_i & q_i \\ p_j & q_j \end{array} \right)=\sum_{j\neq i} (q_i p_j-p_i q_j) \\ \nonumber
=q_i \sum_{j\neq i} p_j - p_i \sum_{j\neq i}  q_j=-q_i p_i+q_i p_i=0
\end{eqnarray}

Thus we see that the theory is anomaly free.

\begin{figure}[h]
\begin{center}
$
\begin{array}{ccc}
{\epsfxsize=2.7cm\epsfbox{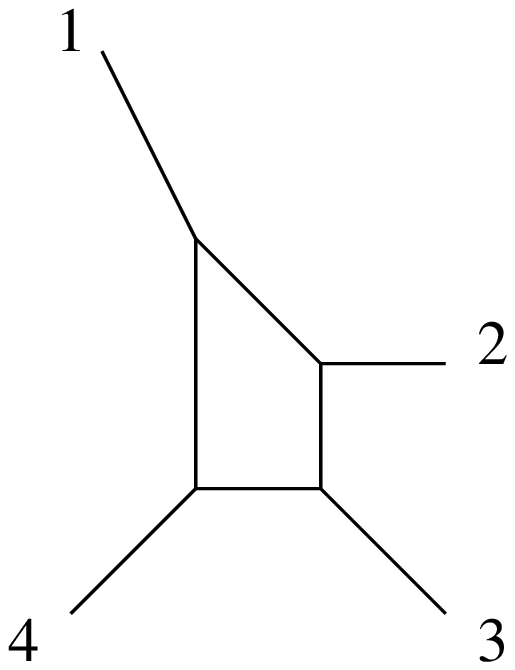}} & \ \ \ \ \ \ \ & {\epsfxsize=3cm\epsfbox{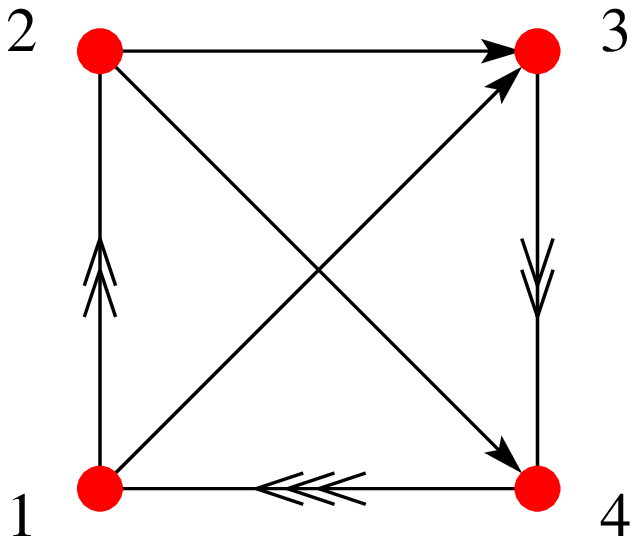}}
\end{array}
$
\end{center}
\caption{A $(p,q)$ web for $dP_1$ and its associated quiver.}
\label{web_to_quiver}
\end{figure}

We conclude this section with an explicit example of how the quiver theory is constructed from the brane web. We consider the case of $dP_1$. A possible
$(p,q)$ web for this geometry is presented in \fref{web_to_quiver}.

We read the $(p,q)$ charges of the 1-cycles from the external legs of the web. They are

\beq
\begin{array}{ccc}
(p_1,q_1)=(-1,2) & \ \ \ \ \ \ \ & (p_2,q_2)=(1,0) \\
(p_3,q_3)=(1,-1) & \ \ \ \ \ \ \ & (p_4,q_4)=(-1,-1) 
\end{array}
\eeq

Using \ref{intersection}, it is immediate to calculate the intersection numbers

\beq
\begin{array}{ccccc}
^\# (C_1.C_2)=-2 & \ \ \ \ & ^\# (C_1.C_3)=-1 & \ \ \ \ & ^\# (C_1.C_4)=3 \\
^\# (C_2.C_3)=-1 & \ \ \ \ & ^\# (C_2.C_4)=-1 & \ \ \ \ & ^\# (C_3.C_4)=-2
\end{array}
\eeq

The resulting quiver is presented in \fref{web_to_quiver}, where we have explicitly labeled the nodes according to the associated
external legs.

\section{Geometric transitions}

In what follows, we will apply the observations in Sections 2 and 3 to obtain the phases of $F_0$ and of all del Pezzo 
surfaces up to $dP_3$. We will use $(p,q)$ web diagrams as representations of the probed geometries and will study
which are the resulting theories after successive blow-ups and blow-downs. This method proves to be very powerful and 
gives all the phases associated to different singularities without doing any calculation!

\subsection{Blow-ups, unhiggsing and $(p,q)$ webs}

\label{geometric}

Del Pezzo surfaces are constructed by blowing-up up to eight generic points on $P^2$. When 
the number of blown up points is less or equal to three, the $SL(3,\IC)$ symmetry of $P^2$ can be used to map the 
generic positions of these points to any desired location. In particular, these can be chosen to be vertices of the web
configuration. Thus, all possible blow-ups of up to three
points can be studied by blowing-up vertices of the web. Using the $(p,q)$ web description of a sphere, the blow-up 
of a point is obtained by replacing it by a segment. The $(p,q)$ charges of the two external legs at the endpoints of the blown-up 2-cycle are given
by the charges of the original external leg. 

The inverse process, a blow-down of a compact 2-cycle to a point, is 
given in the $(p,q)$ web description by the replacement of a segment by a point, and the subsequent combination of the 
external legs attached at the end points of the segment. 
For the four dimensional gauge theory this process is simply a higgsing of 
the $U(1)$ groups associated to both external legs to the linear combination of them under which
the bifundamental field that gets a non-zero vev is neutral.

\begin{figure}[h]
  \epsfxsize = 10cm
  \centerline{\epsfbox{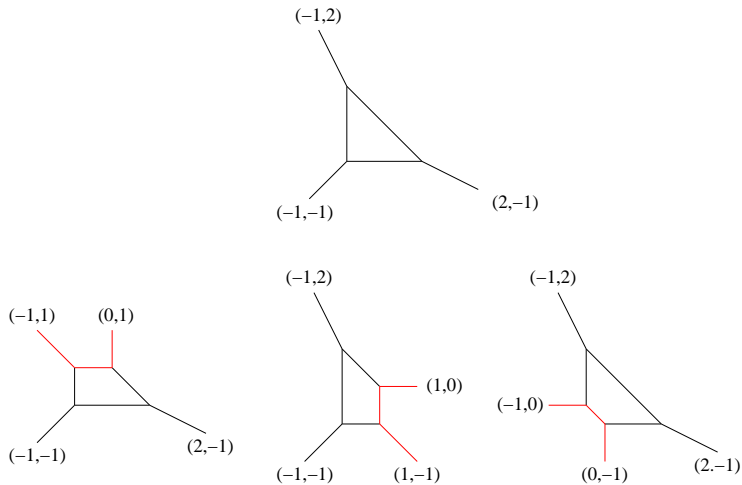}}
  \caption{Possible blow-ups of $dP_0$. All the resulting theories are equivalent.}
  \label{dP1}
\end{figure}

Let us use this method to calculate all the phases associated to D3-branes probing cones over toric del Pezzo surfaces. The starting point is a $(p,q)$ web describing $dP_0$ 
(\fref {dP1}). We have identified
in red the branes obtained as a result of a blow-up. Once we have the resulting webs, we can calculate the intersection matrix and the quiver as in 
\ref{intersection}. The three webs in \fref{dP1} are related by $SL(2,\IZ)$ transformations, and describe the only phase of $dP_1$. 

\begin{figure}[h]
  \epsfxsize = 15cm
  \centerline{\epsfbox{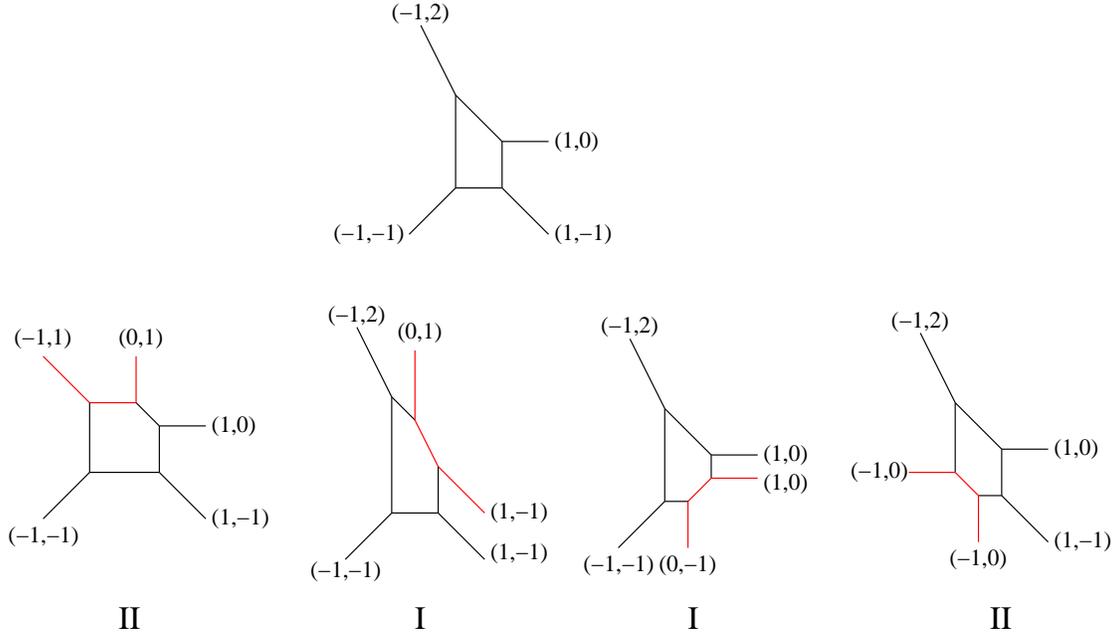}}
  \caption{Possible blow-ups of $dP_1$. They correspond to two inequivalent phases.}
  \label{dP2}
\end{figure}

The following step is taking any of the equivalent webs for $dP_1$ (what we obtain is independent of our choice) and perform all possible blow-ups. 
The results are shown in \fref{dP2}. After calculating the associated quivers we conclude that these webs represent two different theories. 
These are exactly the two toric dual models encountered in \cite{phases} for $dP_2$. At this point a new feature, which will persist for other geometries,
appears: the existence of $(p,q)$ webs with parallel external legs. These models are completely sensible as long as the 4d gauge theories are concerned.
However, these theories present problematic issues when a 5d interpretation of the webs is intended. They may exhibit a leakage of 5d
global charges due to states corresponding to strings stretching between parallel branes, and can also have directions in the moduli space along which the 
superpotential is not convex \cite{webs2}. These drawbacks can be eliminated by providing a suitable UV completion of the theories, embedding them into larger
$(p,q)$ webs. 

\begin{figure}[h]
  \epsfxsize = 13cm
  \centerline{\epsfbox{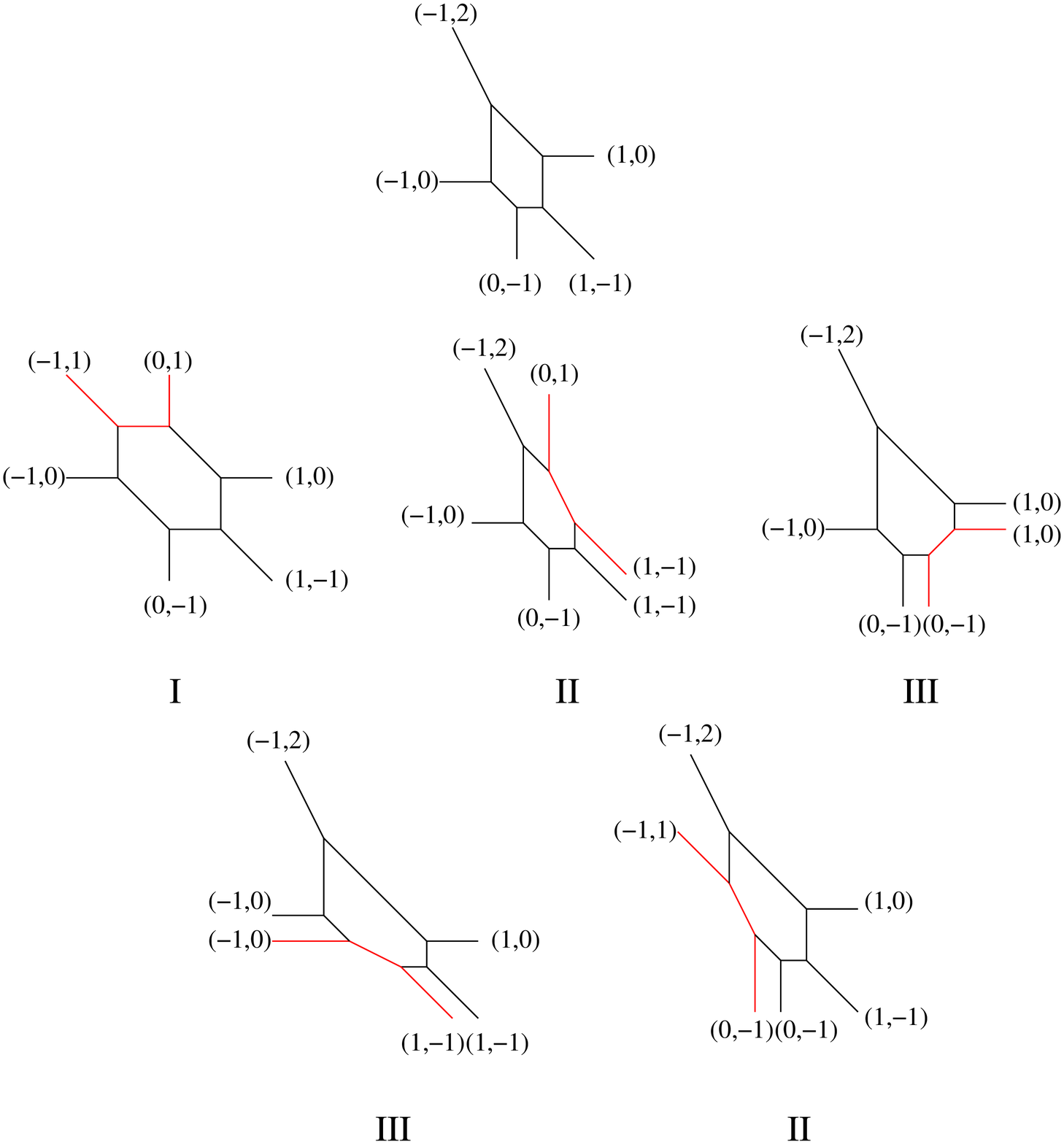}}
  \caption{Possible blow-ups of phase I of $dP_2$.}
  \label{dP3_1}
\end{figure}

\begin{figure}[h]
  \epsfxsize = 13cm
  \centerline{\epsfbox{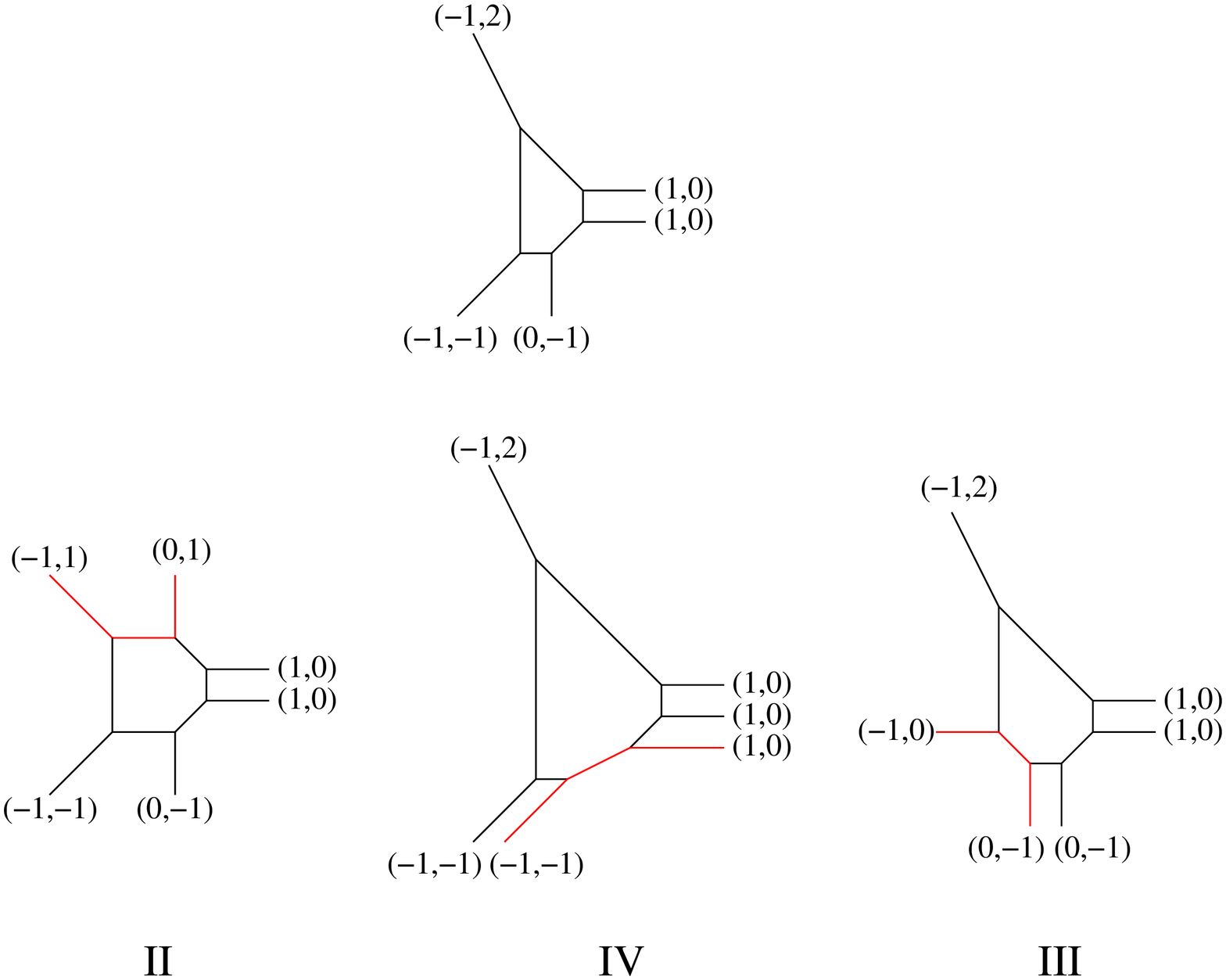}}
  \caption{Possible blow-ups of phase II of $dP_2$.}
  \label{dP3_2}
\end{figure}

We now move on, take one representative for each phase of $dP_2$ and proceed to blow-up points. At this point, it is not possible to blow-up 
any vertex in the web, since in some cases this would lead to crossing external legs (additional compact 4-cycles). The theories coming from the first 
phase and second phase of $dP_2$ are presented in figures \ref{dP3_1} and \ref{dP3_2}. Computing the quivers we obtain the four toric phases 
of $dP_3$ \cite{dual,Chris2}.

\begin{figure}[h]
  \epsfxsize = 12cm
  \centerline{\epsfbox{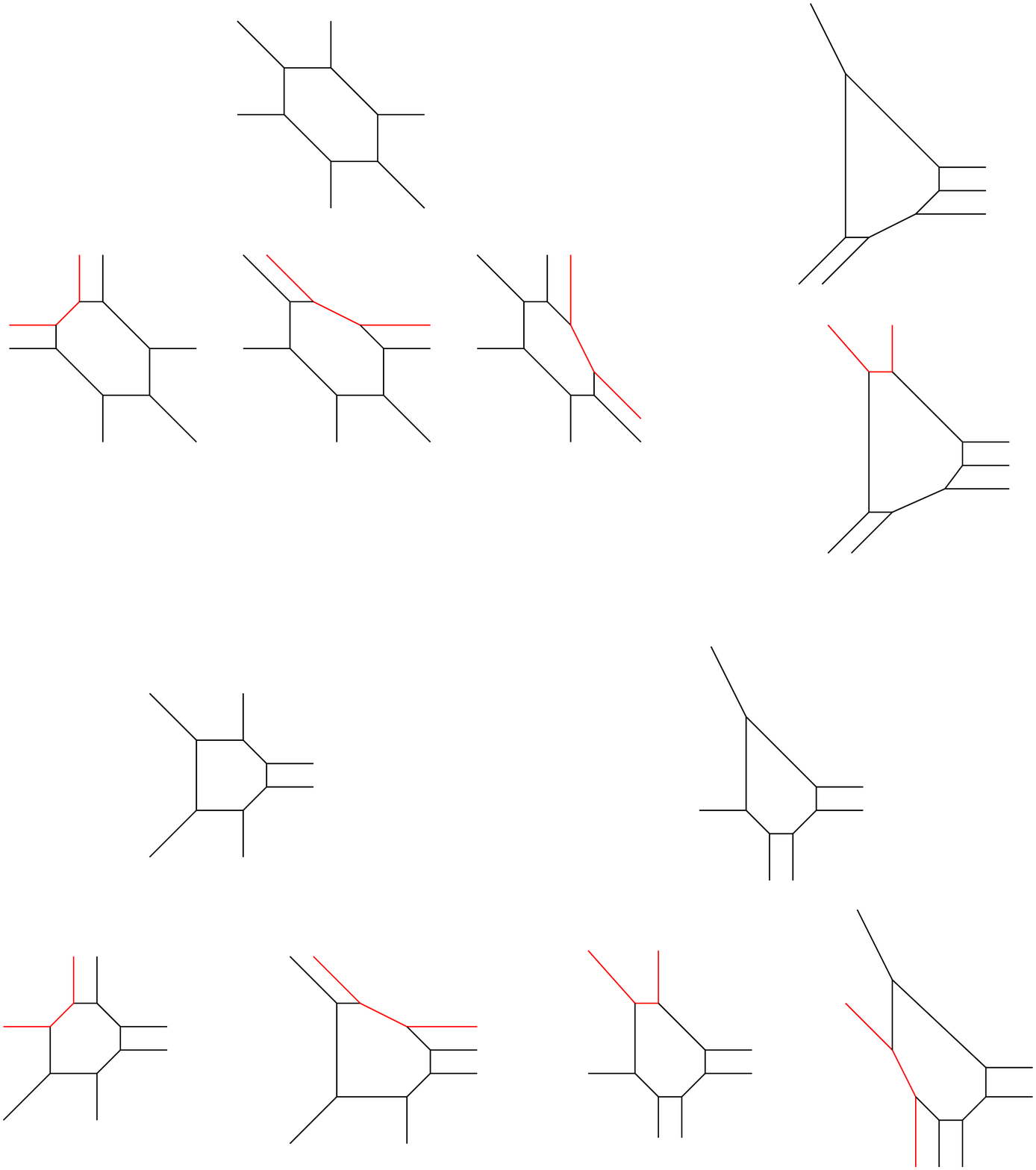}}
  \caption{Possible blow-ups of the four phases of $dP_3$ at non-generic points. We have suppressed $(p,q)$ charges for simplicity.}
  \label{dP4}
\end{figure}

The $SL(3,\IC)$ freedom is exhausted after blowing up three points on $\IP^1$. Thus, we cannot map a further generic point to a vertex of the web and 
then blow it up. This is a manifestation of the fact that $dP_n$ surfaces do not admit a toric description beyond $n=3$. Nevertheless, we can study the 
theories obtained from $dP_3$ after a non-generic toric blow-up. We summarize the possibilities in \fref{dP4}. These $(p,q)$ webs define two quiver 
theories that are studied in detail in \cite{unhiggsing}.

We close this section by emphasizing that different Seiberg dual phases can be understood as related
by blowing-down a 2-cycle and blowing-up a point. This is nothing more than an $SL(3,\IC)$ transformation
relocating one of the blown-up point in $\IP^2$. In this way, the set of Seiberg duality transformations
(that do not change the rank of the gauge groups, keeping them all equal) $G_S$ satisfies

\beq
G_S \subset {SL(3,\IC) \over SL(2,\IZ)}
\eeq

The quotient by $SL(2,\IZ)$ eliminates those transformations that trivially do not change the intersection matrix,
not leading to a dual phase. Furthermore we see that, since there are elements in $SL(3,\IC)$ 
that do not preserve the intersection numbers, some singularities can be represented by 
both $(p,q)$ webs with parallel legs and by $(p,q)$ webs without them.

\subsection{The two phases of $F_0$}

\label{phases_F0}

\begin{figure}[h]
  \epsfxsize = 12cm
  \centerline{\epsfbox{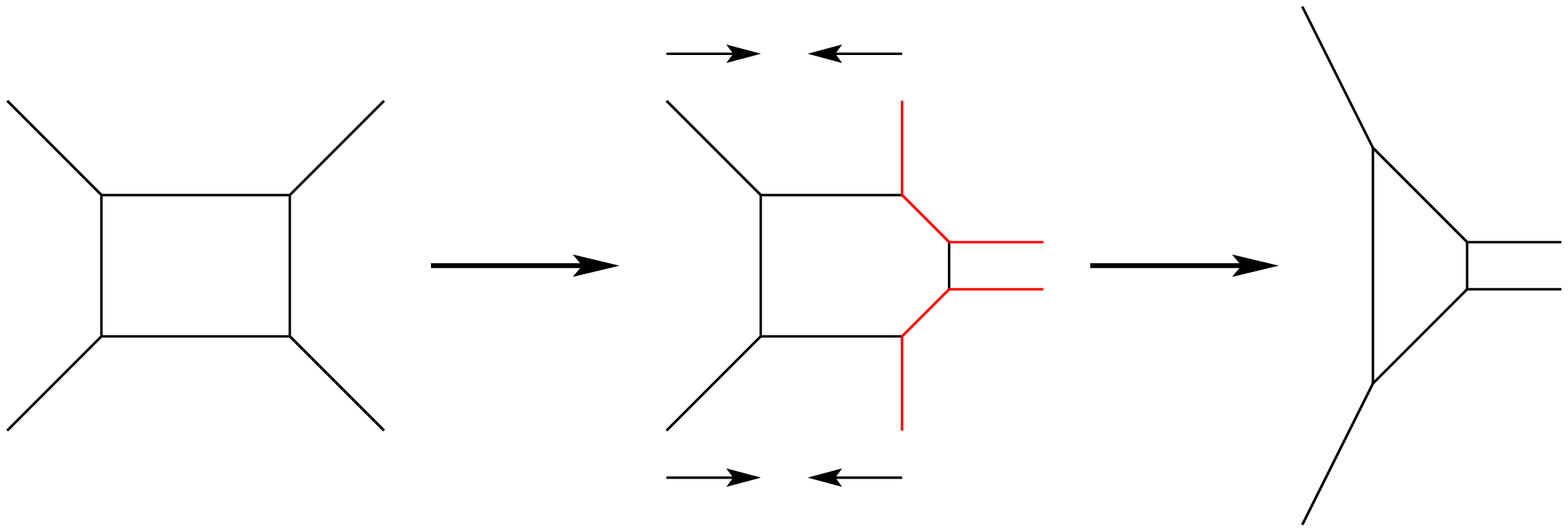}}
  \caption{A possible transition between the two phases of $F_0$, by blowing-up two points and blowing-down two 2-cycles.}
  \label{F01_to_F02}
\end{figure}

The techniques introduced in Section \ref{geometric} can be used to study, for another example, how different phases and singularities are interconnected
by the geometric processes of blowing-up points and shrinking 2-cycles. In \fref{F01_to_F02} we present a possible path
connecting the two phases of $F_0$. Starting from phase I, we blow up two points, arriving at model II of $dP_3$. The last
step in the flow consists of shrinking two 2-cycles to zero size, combining the corresponding external legs.

The intersection matrices for both phases can be computed using \ref{intersection}, and are presented in the appendix.

\subsection{Higgsings as blow-downs}

\label{blowdown}

We have mentioned in Section \ref{geometric} that blow-ups of the geometry correspond to unhiggings when we look at them from the 
perspective of the four dimensional gauge theory on the world volume of the brane probing the singularity. Conversely, 
the blow-down of a compact 2-cycle to a point translates into the higgsing 
of two $U(1)$ factors to a single $U(1)$ by giving a non-zero expectation value to a bifundamental chiral field. 

As we have discussed, compact 2-cycles are represented by the internal finite segments of 
the $(p,q)$ web, their length given by the volume of the corresponding $P^1$'s. Thus, we 
see the beautiful interplay between the two descriptions of the process. When blowing-down, 
we reduce the length of a segment. When this length vanishes, the external legs that are 
located at its endpoints are combined, adding their $(p,q)$ charges. The two 
original $U(1)$'s merge into a single one. The resulting linear combination depends on 
the relation between the coupling constants and is such that a bifundamental field charged under 
the original gauge groups is neutral with respect to it. 

\begin{figure}[h]
  \epsfxsize = 9cm
  \centerline{\epsfbox{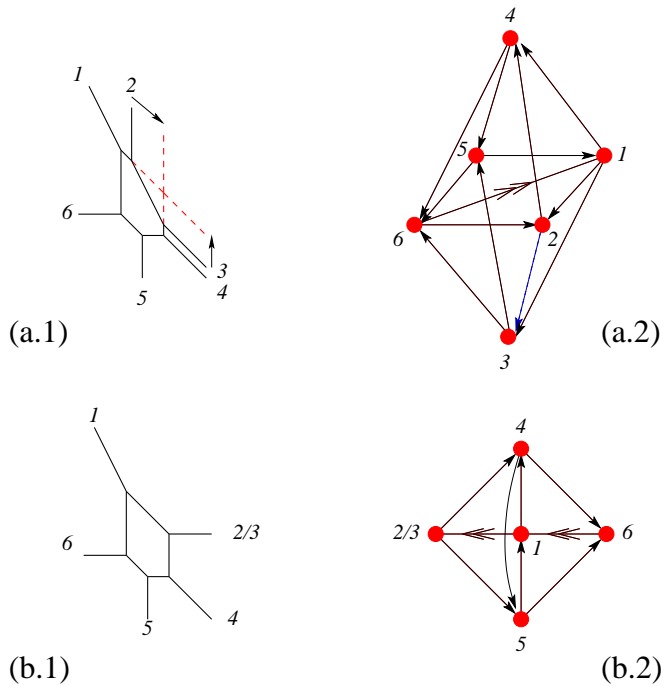}}
  \caption{Higgsing from $dP_3$ (a.1) to $dP_2$ (b.1) by blowing down a 2-cycle. Their 
  corresponding quivers are a.2 and b.2.}
  \label{higgsing}
\end{figure}

The $(p,q)$ description of the original and final geometries allows an immediate identification of which vev we have to 
turn on in the gauge theory in order to flow down to the desired theory. Furthermore, it supplies a correspondence between gauge
groups in the original and final theories. An explicit example of a blow-down from $dP_3$ to 
$dP_2$ will help to clarify these concepts. Let us connect the transition between the theories in \fref{higgsing}. From the respective
webs, we already see that they are related by the combination of nodes 2 and 3.

Before going on, let us notice that we have represented both quivers in a way that makes their symmetries explicit. 
The identification of these symmetries is immediate following the rules that will be presented in Section \ref{symmetries}. Theory A 
has a $\IZ_2 \times \IZ_2$ node symmetry. The first $\IZ_2$ interchanges nodes 3 and 4. The second $\IZ_2$ acts as a 
$\pi$ rotation around the (34) axis and a charge conjugation of all fields. These two symmetries disappear when we combine 2 and 3, 
but a new $\IZ_2$ symmetry that interchanges $2/3 \leftrightarrow 6$, $4 \leftrightarrow 5$ and charge conjugate
all fields appears in model B. At this point we can calculate the intersection matrices and see that theory A
has 14 fields, while theory B has 11. One of the missing fields is the one getting a non-zero vev, so we can already see
that masses for two fields will be generated when higgsing (notice that we know that this mass term will appear without looking at any 
superpotential!) \footnote{It is indeed very simple to understand how these two fields acquire masses. In the brane web language, the combined $2/3$ leg 
is parallel to the existing 6, making the original $\phi_{62}$ and $\phi_{36}$ disappear. This is due to the existence of a $(\phi_{62} \phi_{23} \phi_{36})$ cubic
term in the superpotential, which becomes a mass term after giving a non-zero expectation value to $\phi_{23}$. However, the reader should
be aware that the general situation is that not all gauge invariant operators permitted by a given $(p.q)$ web (alternatively by its associated
quiver) appear in the corresponding superpotential.}.

The original superpotential is \cite{toric,phases}

\begin{eqnarray}
\label{WA}
W_A=\phi_{62} \phi_{24} \phi_{46}+\phi_{51} \phi_{14} \phi_{45}-\phi_{62} \phi_{23} \phi_{36}-\phi_{51} \phi_{13} \phi_{35}+\phi_{13} \phi_{36} \phi_{61} \\ \nonumber
+\phi_{14} \phi_{46} \tilde\phi_{61}+\phi_{23} \phi_{35} \phi_{56} \tilde\phi_{61} \phi_{12}-\phi_{24} \phi_{45} \phi_{56} \phi_{61} \phi_{12}
\end{eqnarray}

The combination of legs 2 and 3 corresponds to turning on a non-zero vev for $\phi_{23}$ in 
\ref{WA}. When doing so, $U(1)^{(2)} \times U(1)^{(3)}$ is broken down to a single $U(1)$, 
under which $\phi_{23}$ is neutral. At the same time two fields, $\phi_{62}$ and $\phi_{36}$, become massive as predicted. We are interested in the IR
limit of this theory, so we integrate them out using their equations of motion. Setting $<\phi_{23}>=1$, the 
resulting superpotential is

\begin{eqnarray}
W=-\phi_{51} \phi_{13} \phi_{35}+\phi_{51} \phi_{14} \phi_{45}+\phi_{14} \phi_{46} \tilde\phi_{61}+ \phi_{35} \phi_{56} \tilde\phi_{61} \phi_{12} \\ \nonumber
+\phi_{13} \phi_{24} \phi_{46} \phi_{61}-\phi_{24} \phi_{45} \phi_{56} \phi_{61} \phi_{12}
\end{eqnarray} 
which is exactly the superpotential of the $dP_2$ theory under consideration \cite{toric,phases}.


\subsubsection{An application, partial resolutions of $\IC^3/(\IZ_3 \times \IZ_3)$}

In Sections \ref{geometric} and \ref{phases_F0}, we obtained all the gauge theories associated to blow-ups of $\IP^2$ and $\IP^1 \times \IP^1$ in a constructive way, identifying
at every step the possible geometric blow-ups. On the other hand, in section \ref{blowdown} we traced the connection between blow-downs, higgsings and transformations
of the $(p,q)$ webs. Let us now consider an example where all these tools and ideas converge.

\begin{figure}
\begin{center}
$
\begin{array}{c}
\begin{array}{c} \mbox{$\IC^3/(\IZ_3 \times \IZ_3)$} \\ {\epsfxsize=4.5cm\epsfbox{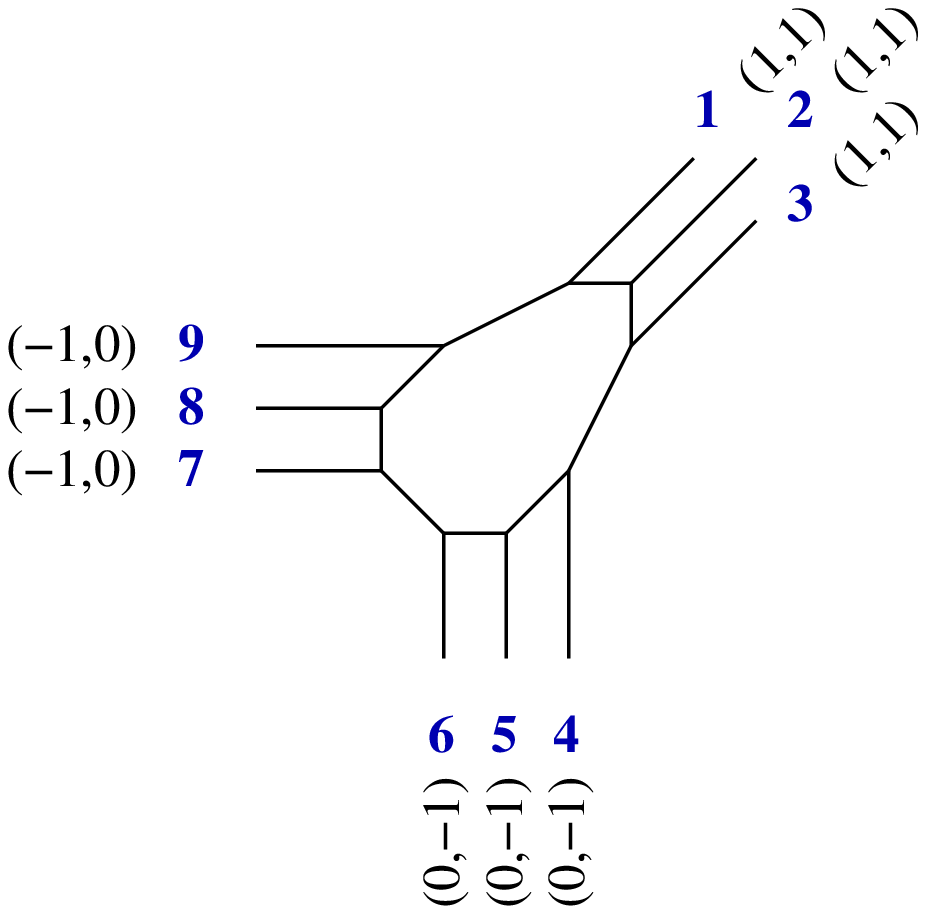}} \end{array} \\

\begin{array}{ccc} \begin{array}{c} \mbox{\underline{Model I}} \\ {\epsfxsize=3cm\epsfbox{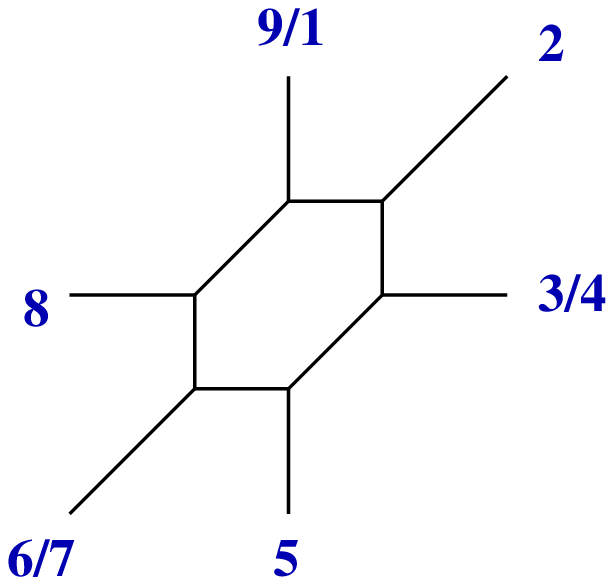}} \\ \phi_{91},  \phi_{34},  \phi_{67} \end{array} &
		   \ \ \ \ \ \ \ \ \ \ \ \ \ &
                  \begin{array}{c} \mbox{\underline{Model II}} \\ {\epsfxsize=4cm\epsfbox{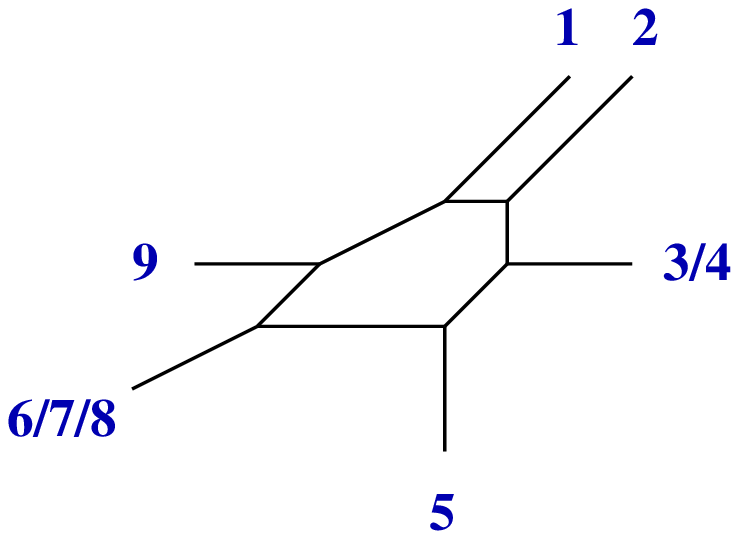}} \\ \phi_{34},  \phi_{67},  \phi_{68} \end{array} \\ \\

                  \begin{array}{c} \mbox{\underline{Model III}} \\ {\epsfxsize=4cm\epsfbox{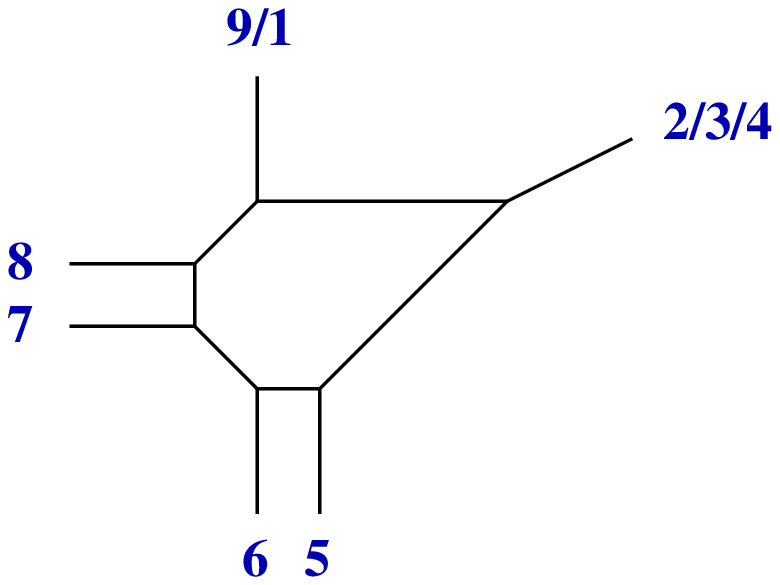}} \\ \phi_{91},  \phi_{24},  \phi_{34} \end{array} &
                  \ \ \ \ \ \ \ \ \ \ \ \ \ &
                  \begin{array}{c} \mbox{\underline{Model IV}} \\ {\epsfxsize=4cm\epsfbox{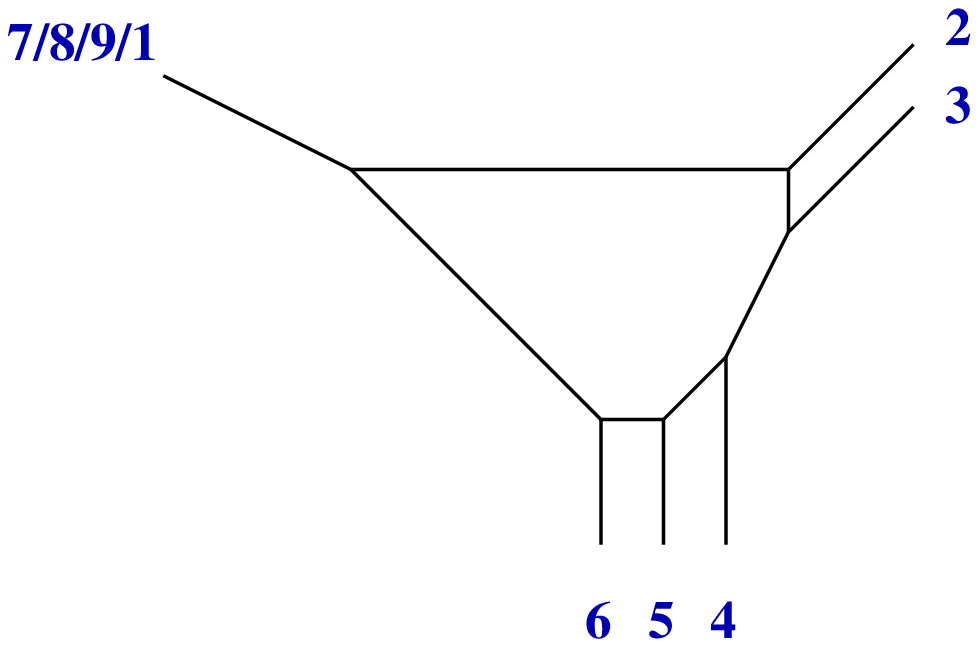}} \\ \phi_{71},  \phi_{81},  \phi_{91} \end{array} 
\end{array}
\end{array}
$
\end{center}
\caption{The four $dP_3$ phases obtained as partial resolutions (higgsings) of $\IC^3/(\IZ_3 \times \IZ_3)$. We indicate
the scalars that get a non-zero vev in each case.}
\label{resolution}
\end{figure}

The four phases of $dP_3$ were presented in Section \ref{geometric}. Furthermore, we have associated specific $(p,q)$ webs to each of them. These theories were obtained 
in \cite{dual,Chris2} by the method of partial resolution of $\IC^3/(\IZ_3 \times \IZ_3)$. Let us see how these results can be recovered immediately using our
techniques. The starting point is the $(p,q)$ web for $\IC^3/(\IZ_3 \times \IZ_3)$ (\fref{resolution}). In each case, we easily see which external legs have to 
be combined in order to get the desired phase. This is not the end of the story, the web construction also tells which fields have to get a non-zero vev in the original theory, they are
the fields associated to the non-vanishing intersections of the combined 2-cycles. We summarize our results 
in \fref{resolution}, indicating the scalars that get a non-zero expectation value.

\section{Symmetries of the four dimensional gauge theory}

\label{symmetries}

An appealing feature of the $(p,q)$ language is that it makes quiver symmetries of the gauge theory evident. We will 
consider here two examples of how this symmetries manifest in the brane representation. These symmetries have
been studied in \cite{Chris2,symmetries}, along with their importance as a tool for determining the structure of superpotentials.

\bigskip

\begin{figure}[h]
  \epsfxsize = 9cm
  \centerline{\epsfbox{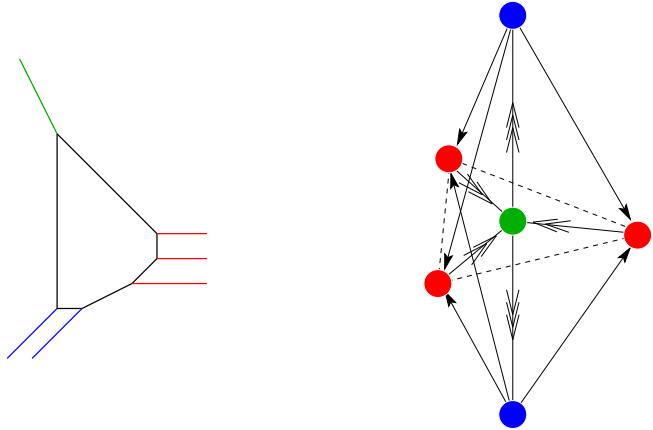}}
  \caption{$(p,q)$ web for phase IV of $dP_3$. We have colored the external branes giving rise to the different 
           $U(1)$ factors accordingly to their transformation properties under quiver symmetries.}
  \label{global_syms1}
\end{figure}

\underline{$S_n$ symmetries}: These symmetries appear when the web brane configuration has sets of 
$n$ parallel external legs (in the geometric language non-compact 2-cycles with the same $(p,q)$ charges). Parallel 
branes have vanishing mutual intersections, while their intersections with the rest of the branes are identical. Due 
to this identity of the intersections, the gauge groups associated to parallel legs can be permuted leaving the quiver 
invariant. The symmetry group in this case is the full $S_n$ permutation group. These discrete symmetries get 
enhanced to a continuous $SU(n)$ when the parallel branes coincide. In \fref{global_syms1} we show phase IV 
of $dP_3$ as an example. The three parallel red legs give rise to a $S_3$ symmetry between red nodes in the quiver, while
there is a $\IZ_2$ symmetry that interchanges blue nodes coming from the parallel blue branes.

\bigskip

\begin{figure}[h]
  \epsfxsize = 7cm
  \centerline{\epsfbox{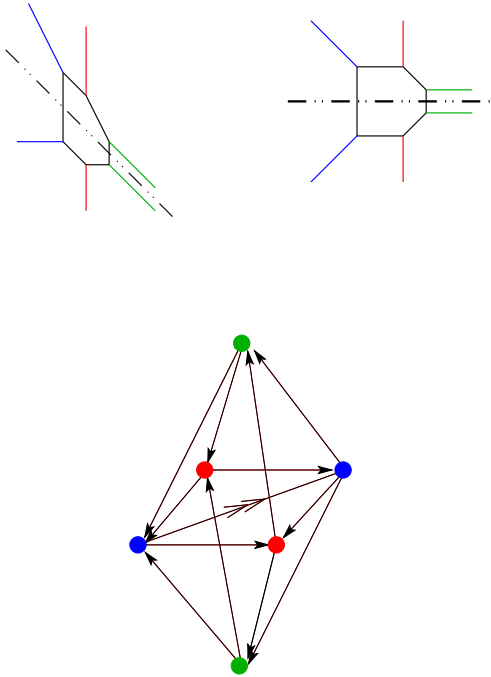}}
  \caption{Two different $(p,q)$ webs for phase II of $dP_3$. Once again, nodes of the same color transform into one another by the quiver symmetries.}
  \label{global_syms2}
\end{figure}

\underline{$\IZ_2$ axial symmetries}: this is another example of symmetries that can be read directly from the $(p,q)$ webs.
They appear whenever the web has an axis of symmetry. In these cases the theories are invariant under exchange of 
gauge groups associated to external legs at both sides of the axis and charge conjugation of all fields. Is it important
to notice that these reflections are indeed a subset of a larger set of transformations, given by all the $GL(2,\IZ)$ symmetries that
map the webs onto themselves (i.e. that preserve the $(p,q)$ charges of external legs). We have chosen to discuss $\IZ_2$ reflections
because they are the simplest of these symmetries, but we will also be considering the case of rotations later in this section. As an 
example, we present phase II of $dP_3$ in \fref{global_syms2}. The two webs in \fref{global_syms2} are equivalent, and correspond to the 
same phase. We have presented both in order to illustrate how the same symmetry can arise in differently looking webs. 
Moreover, this example illustrates how the symmetry axis can sometimes be hidden. As it can be seen in the example,
the axis can be made evident by an $SL(2,\IZ)$ change of $(p,q)$ basis, which preserves the intersection numbers between cycles and simply 
corresponds to a variation in the complex scalar $\tau$ in \ref{tension}. 

For this example,

\beq
\begin{array}{ccc}
e_1=(1,0) & \ \ \ \ \ \ \ & e'_1=(1,0) \\
e_2=(0,1) & \ \ \ \ \ \ \ & e'_2=(1,1)
\end{array}
\eeq

The two bases are related by the $SL(2,\IZ)$ matrix

\beq
C=\left( \begin{array}{cc} 1 & 1 \\ 0 & 1 \end{array} \right)
\eeq

Based on the preceding observations, we can use the $(p,q)$ webs listed in the appendix and make an immediate classification of the mentioned node symmetries 
that appear in each model. The results are summarized in Table \ref{table_symmetries}. 

Three of the models deserve a more detailed explanation. The first of them is $dP_0$, whose $(p,q)$ web has an obvious 
$\IZ_2$ axis of symmetry going along one of its legs. Furthermore, the three external legs are equivalent under $SL(2,\IZ)$
transformations that ``rotate'' the web. As a result, the full node symmetry group of $dP_0$ is $D_3$. An identical reasoning
applies to the first phase of $dP_3$, which has an evident $\IZ_2$ axis, and whose six external legs are equivalent under
$SL(2,\IZ)$, leading to a $D_6$ symmetry. Finally, the fourth phase of $dP_3$ has one set of two and another one of three parallel
external legs. According to our rules, this corresponds to a $\IZ_2 \times S_3=D_6$ symmetry.

\begin{table}
\begin{center}
$
\begin{array}{|c|c|c|}
\hline
        \ \ \ \ \ Singularity  \ \ \ \ \ & \ \ \ \ \ Phase \ \ \ \ \ &  \ \ \ \ \ Node \ Symmetry  \ \ \ \ \ \\ 
        \hline
        \begin{array}{c}  F_0 \\ \\ \end{array}  &  \begin{array}{c}
        \mbox{I} \\ \mbox{II} \end{array}  &  \begin{array}{c} \IZ_2 \times \IZ_2 \\  \IZ_2 \times \IZ_2 \end{array} \\  
        \hline 
        dP_0 & & D_3  \\
        \hline
        dP_1 & & \IZ_2  \\
        \hline
	 \begin{array}{c}  dP_2 \\ \\ \end{array}  &  \begin{array}{c}
        \mbox{I} \\ \mbox{II} \end{array}  &  \begin{array}{c} \IZ_2 \\ \IZ_2 \end{array} \\ 
        \hline 
         \begin{array}{c}  dP_3 \\ \\ \\ \\ \end{array}  &
        \begin{array}{c} \mbox{I} \\ \mbox{II} \\ \mbox{III} \\ \mbox{IV} \end{array}  &
        \begin{array}{c} D_6 \\  \IZ_2 \times \IZ_2 \\ \IZ_2 \times \IZ_2  \\ D_6  \end{array} \\ 
        \hline    
\end{array}
$
\end{center}
\caption{Classification of node symmetries.}
\label{table_symmetries}
\end{table}

\section{Geometric transitions from the perspective of five dimensional theories}

As we have discussed, $(p,q)$ webs give a representation of the moduli space of ${\cal N}=1$ theories in 4 dimensions, 
for those cases in which the moduli space admits a toric description. For these models, brane webs are simply the toric skeletons 
representing vanishing cycles. At the same time, $(p,q)$ webs can be used to study 5 dimensional ${\cal N}=1$ gauge theories living
in the $4+1$ common dimensions of the branes. The purpose of this section will be to understand the translation of the four dimensional
concepts of Seiberg duality and of different phases, to the five dimensional language. While doing so, we will get some nice 
dynamical information about the five dimensional theories.


\subsection{Five dimensional interpretation of the theories}

We have already discussed how $(p,q)$ webs lead to 5 dimensional $SU(N_c)$ theories with $N_F$ flavors. The number of
colors is given by the number of parallel internal branes. For all the cases we are studying, the $(p,q)$ webs posses only one 
closed face (i.e. a single compact 4-cycle in the geometric interpretation), and have a pair of parallel internal
branes, so they will be associated to $SU(2)$ theories. In all the webs sketched in Figures \ref{dP1} to 
\ref{F01_to_F02}, it is possible to identify at least one such a pair of parallel finite branes that play the role of color branes. 

Of the $N_L$ external legs of a web, four have to be the supporting structure of color branes. It is also possible
to see that in all the studied webs, after we identify the supporting branes, the remaining ones result to be
parallel to color branes, thus admitting an interpretation as $N_F=N_L-4$ flavor branes. Putting all these
things together we see that $F_0$ phases will be associated to $SU(2)$ with no flavors, while $dP_n$ theories will be
represented in five dimensions by $SU(2)$ models with $N_F=n-1$.

\begin{figure}[h]
  \epsfxsize = 8cm
  \centerline{\epsfbox{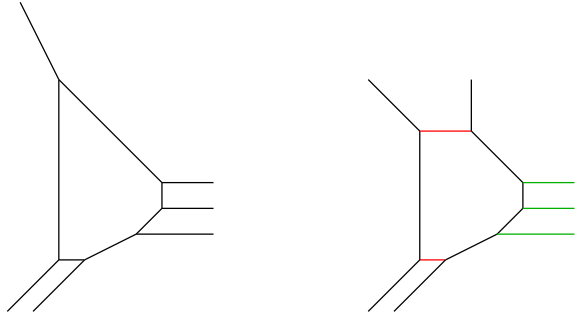}}
  \caption{Phase IV of $dP_3$. It can be interpreted in five dimensions as $SU(2)$ with 3-1 flavors.}
  \label{3-1}
\end{figure}

There are two theories that require a more careful interpretation. The first one is the well-known case of $dP_0$. Having only three external legs, this
theory is understood as $SU(2)$ with $-1$ flavors. We can extend the reasoning in the following way, any time we face a theory where we cannot identify
$N_L-4$ legs as flavors, we blow up $N$ points until reaching a model with the usual interpretation. This theory will have $N_L+N-4$ flavors. Then we say
that the number of flavors in the original theory is $(N_L+N-4)-N$. The other special case is phase IV of $dP_3$, which is shown in \fref{3-1} together with
the same $(p,q)$ web blown-up at one point. According to our previous statements, the five dimensional interpretation of this model is $SU(2)$ with $3-1$ flavors.

\subsection{Different limits in moduli space, a first example}

Let us start studying how the flow between theories can be interpreted as adding a flavor to the corresponding five dimensional gauge theory 
and considering different, eventually infinite, limits in parameter space. To do so, let us focus on the example of a transition between
one of the phases of $F_0$ and $dP_1$. The main point here is to realize that in both brane configurations one of the external legs can be understood
as coming from a junction between two branes, one of which is a flavor brane. When taking the location of this junction very far away from the core of the 
$(p,q)$ web, the configurations become those studied in Sections \ref{geometric} and \ref{phases_F0}. The position of the flavor brane is parametrized by the bare mass of the quark.
When it becomes infinite, the quark decouples leaving us with pure $SU(2)$ theories with no flavors. 

\begin{figure}[h]
  \epsfxsize = 8cm
  \centerline{\epsfbox{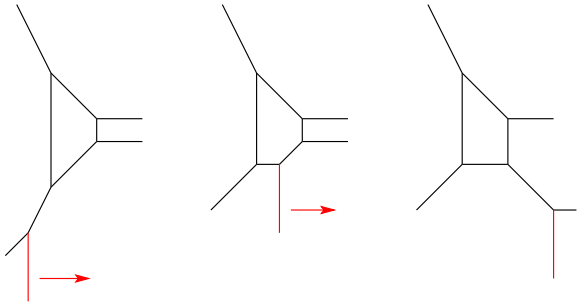}}
  \caption{Flowing from $F_0$ to $dP_1$ by tuning the bare mass of the quark from $m\rightarrow -\infty$ to
           $m\rightarrow \infty$.}
  \label{F0_to_dP1}
\end{figure}

\fref{F0_to_dP1} shows how in the limit $m\rightarrow -\infty$ we have the second phase of $F_0$, for $|m|<\phi/2$ we get phase I of $dP_2$ and finally we obtain
$dP_1$ for $m\rightarrow \infty$.

\subsection{Geometrical blow-ups as tuning bare masses}

Encouraged by the example presented in the previous section, we can ask whether this is a general feature and we may indeed
interpret all geometrical transitions of the type we are considering as tuning the bare mass for some quark. After 
inspecting Figures \ref{dP1} to \ref{F01_to_F02} we conclude that this in fact is true!

The way of seeing this is that, in all cases, one of the two external legs connected to a 2-cycle coming from a blown-up point is parallel
to a pair of finite segments in the inner face of the $(p,q)$ webs. Thus, this external leg can be understood as a flavor brane, while
the two finite branes play the role of $SU(2)$ color branes. We can think about the blowing-up process as bringing the flavor brane from infinity 
($m\rightarrow \pm \infty$) until it reaches the body of the web. We can repeat this process indefinitely. One important point that has
to be kept in mind is that, after each step, the two branes acting as color branes can change (and thus the orientation of the flavor brane we have
to consider).

\begin{figure}[h]
  \epsfxsize = 8cm
  \centerline{\epsfbox{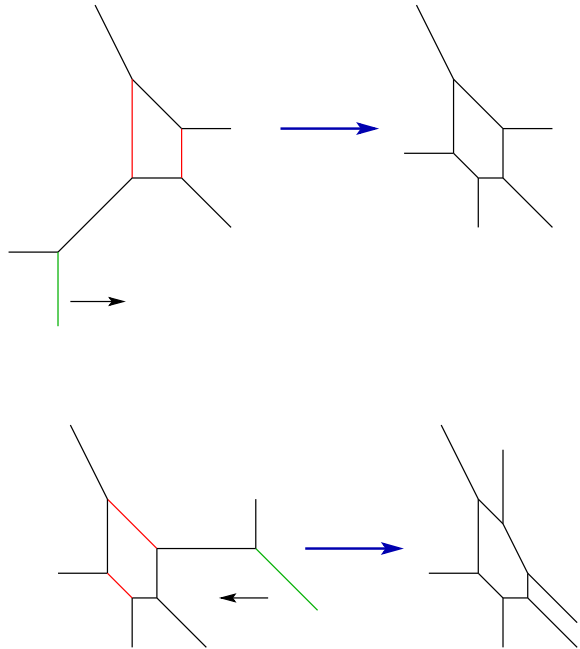}}
  \caption{Flow $dP_1 \rightarrow dP_2 \ II \rightarrow dP_3 \ II$, obtained by bringing bare masses to finite values.}
  \label{tuning_mass}
\end{figure}

\fref{tuning_mass} shows an example of a flow $dP_1 \rightarrow dP_2 \ II \rightarrow dP_3 \ II$, illustrating how the direction of color and flavor branes
can change at each step. As an aside, this example also shows an interesting situation, the fact that at some point in the blow-up process there can be more
than one possible choices of which branes to consider color branes. We see that, before the second blow-up, another legitimate choice would have been the two vertical
finite branes.

\subsection{BPS spectrum}

\begin{figure}[h]
  \epsfxsize = 9cm
  \centerline{\epsfbox{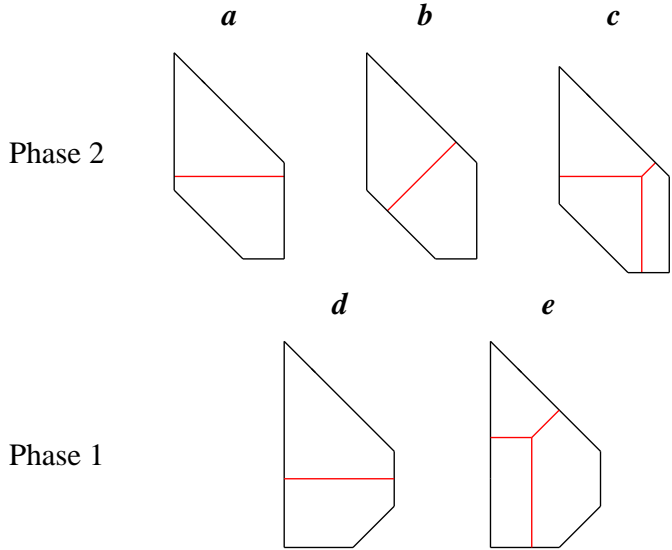}}
  \caption{Some BPS states for phases I and II of $dP_2$. String junctions are represented in red. For simplicity, we have suppressed the external legs of the
$(p,q)$ webs.}
  \label{BPS_dP2}
\end{figure}

BPS states in the five dimensional theory are given by webs of strings ending on the 5-brane 
web \cite{webs2}. We will use this construction to see how
BPS spectra of different phases are related. Let us consider the two phases of $dP_2$ since 
they constitute one of the simplest examples. We will also 
restrict our analysis to BPS states associated to string webs with only two and three end 
points (the extension to other states is immediate). The corresponding
configurations are shown in \fref{BPS_dP2}.

Both spectra are quite different. Specifically there is no analog of state b of phase II in 
phase I. Nevertheless, we have seen that both the
geometric picture and the five dimensional one in terms of varying parameters suggest that 
the passage between different phases is a continuous process.

\begin{figure}[h]
  \epsfxsize = 10cm
  \centerline{\epsfbox{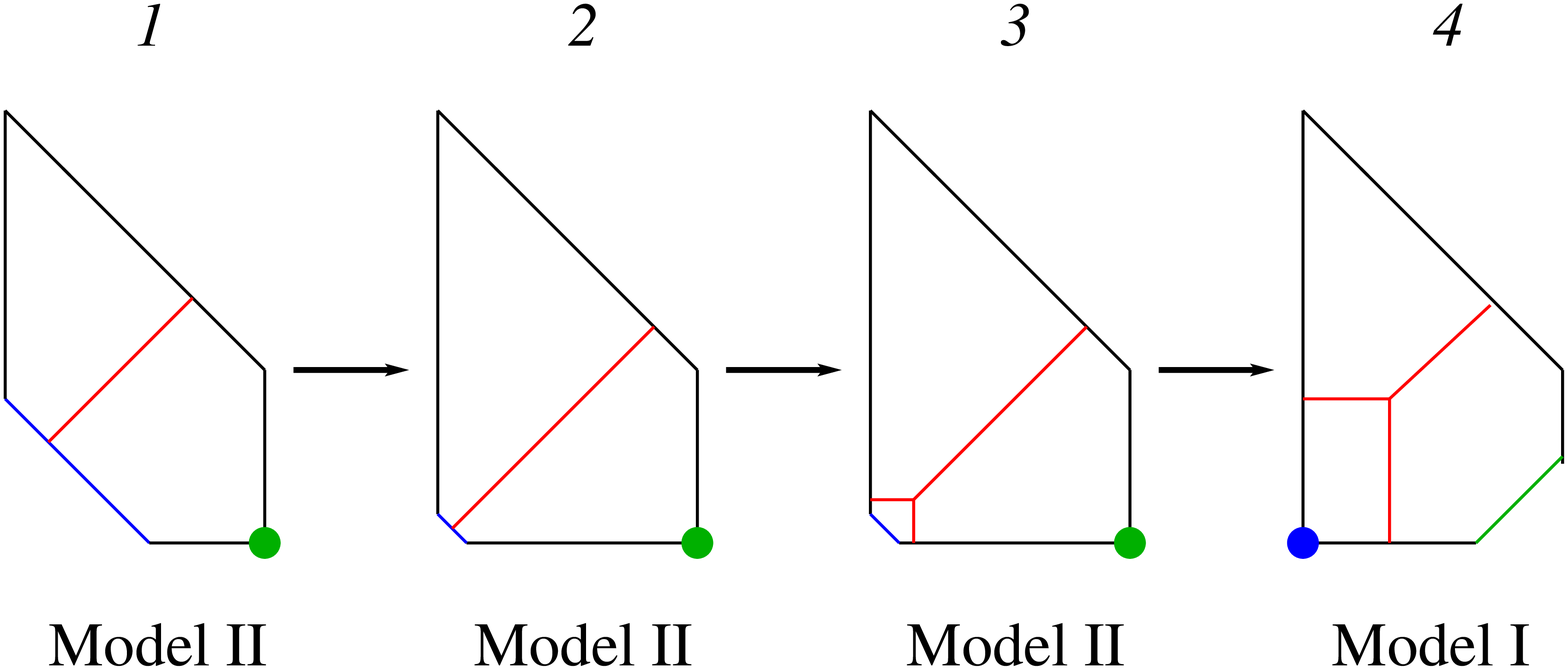}}
  \caption{Continuous flow between BPS states of the two phases of $dP_2$.}
  \label{BPS_flow}
\end{figure}

Let us understand how the two spectra are continuously connected. To do so, we follow the fate 
of state b as we flow between phases II and I. In \fref{BPS_flow} 
we show different stages of this transition. The starting point is phase II and we gradually 
reduce the size of the blue 2-cycle. From stages 2 and 3 we see that,
as the blue cycle approaches zero size, state b becomes degenerate with one state of type c 
(both of them in the BPS spectrum of phase II). State c survives the 
transition, becoming state e of phase I. The final step consists on shrinking the blue 2-cycle 
to a point and blowing up the green point. In conclusion state
b undergoes a continuous transformation into state e, going through a point at which both 
states have the same mass.

We have understood that, for the specific case of geometric transitions between $dP_2$ phases, 
BPS states corresponding to string junctions with support on 
collapsing 2-cycles cease to exist as these 2-cycles shrink to a point, but become degenerate 
with other BPS states that remain in the spectrum. The 
same conclusion can be reached in the general case. 

\begin{figure}[h]
  \epsfxsize = 7cm
  \centerline{\epsfbox{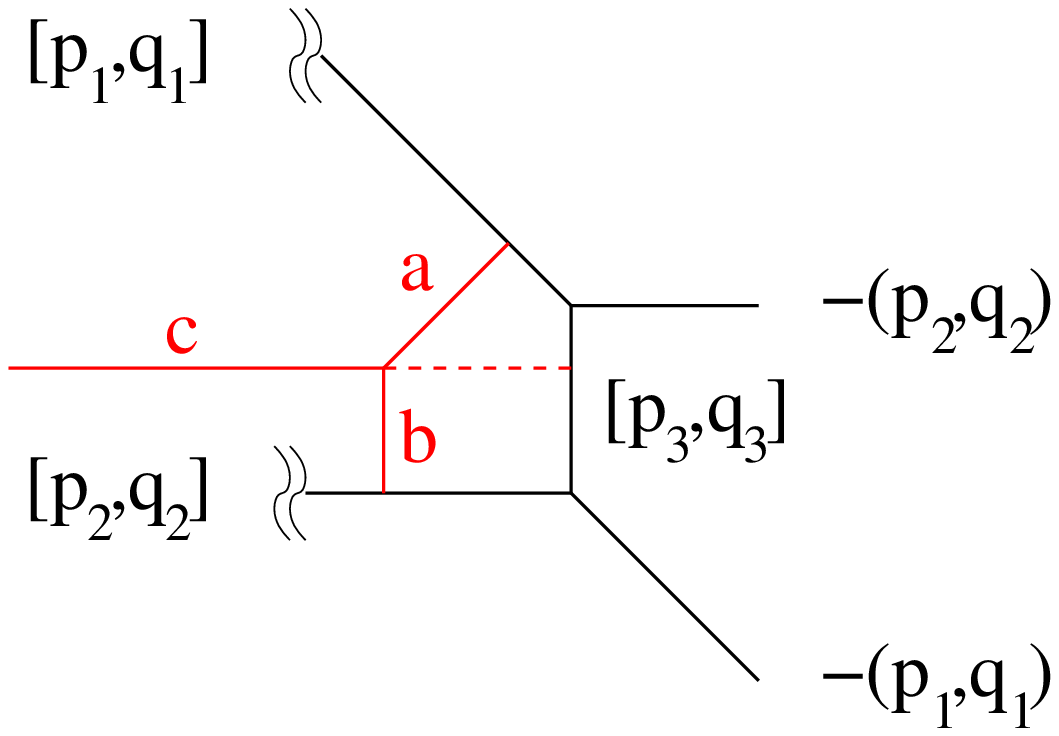}}
  \caption{Continuous connection between BPS states in the general case.}
  \label{BPS_general}
\end{figure}

The string junctions (BPS states in five dimensions) that can in principle disappear abruptly are those ending on shrinking 
blown-up 2-cycles (exceptional curves). It is easy to see (and Figures \ref{dP1} to \ref{F01_to_F02} illustrate this fact) that external legs attached to exceptional curves
are parallel to internal branes ending on the corresponding segment of the $(p,q)$ web. The general situation is presented in \fref{BPS_general}, where
the charges of internal branes are shown between brackets, and those of external legs 
between parentheses.

$(p,q)$ charge conservation fixes the slope of the shrinking brane to be given by

\beq
[p_3,q_3]=[p_1-p_2,q_1-q_2]
\eeq

Let us now consider the string junction shown in red in \fref{BPS_general}, which can be part of a larger BPS configuration. Its a and b legs are
perpendicular to branes 1 and 2. Thus

\beq
\begin{array}{c}
(p_a,q_a)=(q_1,-p_1)\\
(p_b,q_b)=(-q_2, p_2)
\end{array}
\eeq

Once again, $(p,q)$ charge conservation implies

\beq
(p_c,q_c)=(q_1-q_2,-p_1+p_2)=(q_3,-p_3)
\eeq

Then, string c is perpendicular to brane 3. We can now follow a reasoning identical to the one used previously for $dP_2$. As brane 3
goes to zero size, legs a and b of the string junction can be negligible small, and the corresponding BPS state becomes degenerate with the 
one ending directly on brane 3. This concludes our proof of the continuity of BPS spectra in the general case.

We discussed in Section \ref{geometric} how Seiberg dual theories are related by the combination
a blow-down and a blow-up in the non-compact Calabi-Yau probed by the D3-brane. The arguments
presented in this section lead us to an important conclusion, Seiberg duality between theories 
in four dimensions, appears in the associated five dimensional models 
as crossing curves of marginal stability.

\subsection{Continuity of the monopole tension}

In this section we will calculate another five dimensional quantity, the monopole tension, and study how its values for
different theories are related

Let us consider the example of the two phases of $dP_2$, which in five dimensional language correspond to $SU(2)$ theories with one flavor. Calculating the monopole tensions for these 
two phases (which is simply the area of the inner face of the $(p,q)$ web \cite{webs2}) we find

\begin{eqnarray}
T_M^{(I)}&=&{7\over 8} \phi^2+ \left( {1 \over g_0^2}+{m\over 2} \right) \phi-{m^2\over 2} \\ \nonumber
T_M^{(II)}&=&{7\over 8} \phi^2+ {\phi \over g_0^2}-{m^2\over 2} 
\end{eqnarray}

We see that this quantity has a different functional dependence on the parameters when we 
consider the two theories. Another question (whose answer is obvious from
the $(p,q)$ web picture) is what the value of $T_M^{(I)}$ is when $m=\phi/2$. In this situation we know that one of the quarks becomes massless, while the other one
gets the same mass that the gauge boson $m_Q=\phi$. The tension in this case becomes

\beq
T_M=\phi^2+{\phi \over g_0^2}
\eeq
which is exactly that for $dP_1$. Once again, the transition is continuous.

\section{Conclusions}

In this paper we have studied dualities and flows between gauge theories living on
D3-branes probing toric singularities. We have found $(p,q)$ webs very useful for this task, and for
establishing relations among the probed geometry, the four dimensional theories on the world volume of the branes and 
five dimensional $SU(2)$ associated theories.

In Section 4 we have interpreted the flow between the four dimensional theories corresponding
to the zeroth Hirzebruch and the del Pezzo surfaces as geometric transitions in the probed singularities. 
We also established the geometric transformations connecting toric dual models. In doing so, the $(p,q)$ web 
representation of the toric varieties became not only a useful pictorial representation of the process, but
a whole computational tool. The process of obtaining all the theories is reduced to doing successive blow-ups
and to calculate intersection matrices. This simplicity can be contrasted with methods previously
employed for the same task, based on partial resolutions of $\IC^3/(\IZ^3 \times \IZ^3)$, which are
computationally much more involved. Another advantage of the $(p,q)$ web approach is that it offers
a geometric intuition at every point of the process. 

Furthermore, we studied the connection between blow-downs(ups) and (un)higgsings in the four dimensional theories.
When doing so, the associated $(p,q)$ webs permit the immediate identification of which field must acquire
a non-zero vev. This was exemplified by getting the partial resolutions of $\IC^3/(\IZ^3 \times \IZ^3)$ that
give the four toric dual theories corresponding to $dP_3$. 

Section 5 was devoted to study how quiver symmetries can be read off from $(p,q)$ webs. 
The web constructions make these symmetries evident. The identification of the symmetry groups is reduced
to counting parallel external legs and finding axes of symmetry.

In Section 6 we initiated the exploration of a new perspective for geometric transitions. Exploiting the connection provided
by $(p,q)$ webs, we developed the interpretation of the 
studied theories as five dimensional $SU(2)$ gauge theories with $N_F$ flavors. We showed how geometrical blow-ups can 
be understood as bringing flavors from infinite bare mass. We proved that BPS spectra of two theories connected
by a geometric transition are continuously connected. In this language, the transition corresponds to crossing a curve of marginal
stability. We also studied the continuous relation between the monopole tensions in two such theories.

\section*{Acknowledgements}

The authors would like to thank Bo Feng, Yang-Hui He and Amer Iqbal for valuable 
discussions. A.H. would like to thank the organizers of the "M Theory" workshop in
the "Isaac Newton Institute for Mathematical Sciences" for their hospitality while this
work was being completed. S.F. would also like to thank Martin Schvellinger for helpful
conversations. Research supported in part by the CTP and the LNS
of MIT and the U.S. Department of Energy under cooperative agreement
$\#$DE-FC02-94ER40818. A. H. is also supported by the Reed Fund Award and 
a DOE OJI award.

\bigskip

\section{Appendix: Gauge theories for branes on toric singularities}

In this appendix we summarize the theories studied throughout the paper. For each of them we give a 
$(p,q)$ web and its quiver \footnote{In some of the quivers we have charge conjugated all the
fields in order to follow the ones presented in the references \cite{toric,phases,dual,symmetries}.}. 
For the del Pezzo surfaces, we also include the corresponding 
fractional brane charges. These charges were calculated 
with the procedure described in \cite{HI}, which uses the map between 3-cycles in the mirror
manifold and vector bundles on the del Pezzo surfaces of \cite{mirror2,mirror3}. The
purpose of their inclusion is to exemplify how the combination and splitting of external legs of the $(p,q)$ webs are associated
to the same operations on the fractional brane charges.

\bigskip

\newpage

\underline{Cone over $F_0$}

\medskip

$F_0$ has two phases. Phase I has 12 fields, while Phase II has 8.

\beq
\begin{array}{|c|c|c|}
\hline
(p,q) \ web  &  Quiver & Intersection \ matrix \\
\hline 
\ba{c} {\epsfxsize=2.7cm\epsfbox{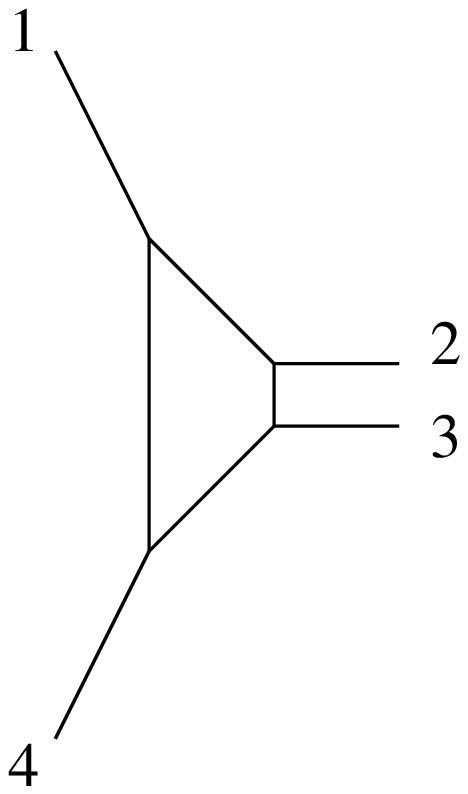}} \ea  & \ba{c} {\epsfxsize=3.2cm\epsfbox{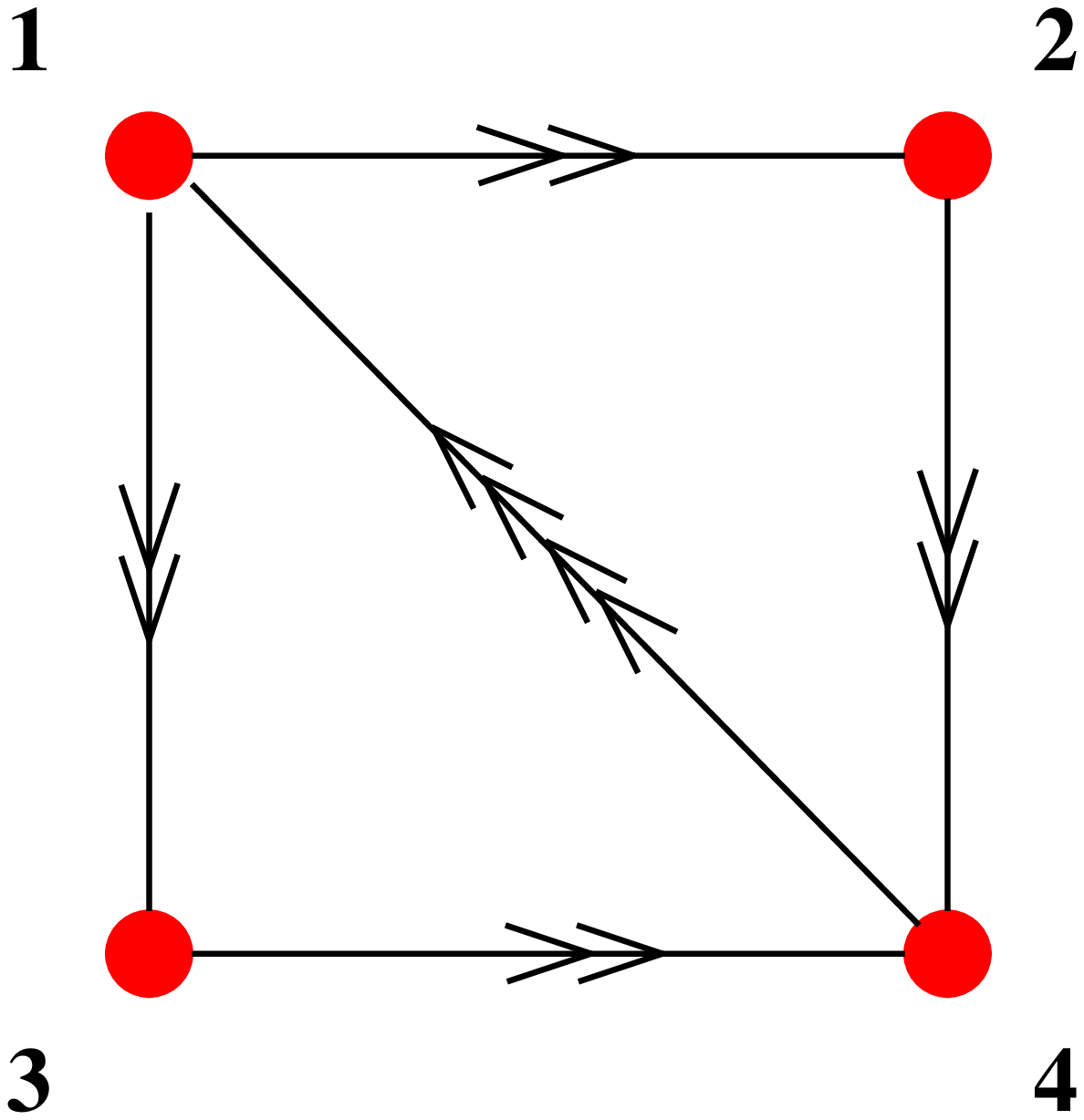}} \ea  & \ba{c} {\cal I}_{I}=\left(\begin{array}{cccc} 0 & -2 & -2 & 4 \\ 2 & 0 & 0 & -2 \\ 2 & 0 & 0 & -2 \\ -4 & 2 & 2 & 0  \end{array} \right) \ea \\

\hline
\ba{c} {\epsfxsize=3.5cm\epsfbox{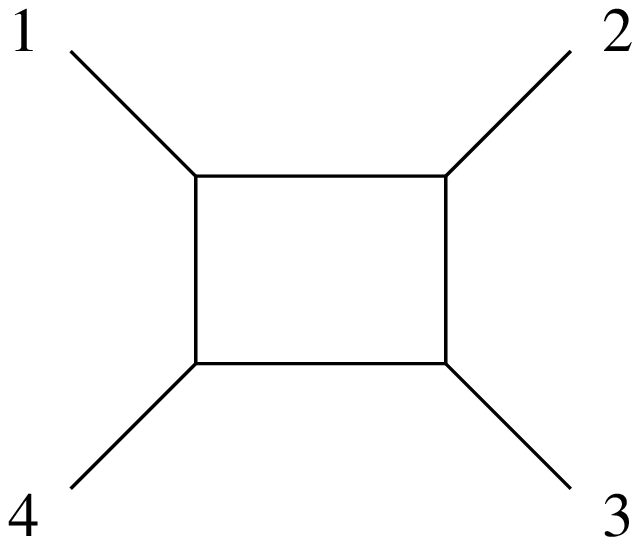}} \ea & \ba{c} {\epsfxsize=3.2cm\epsfbox{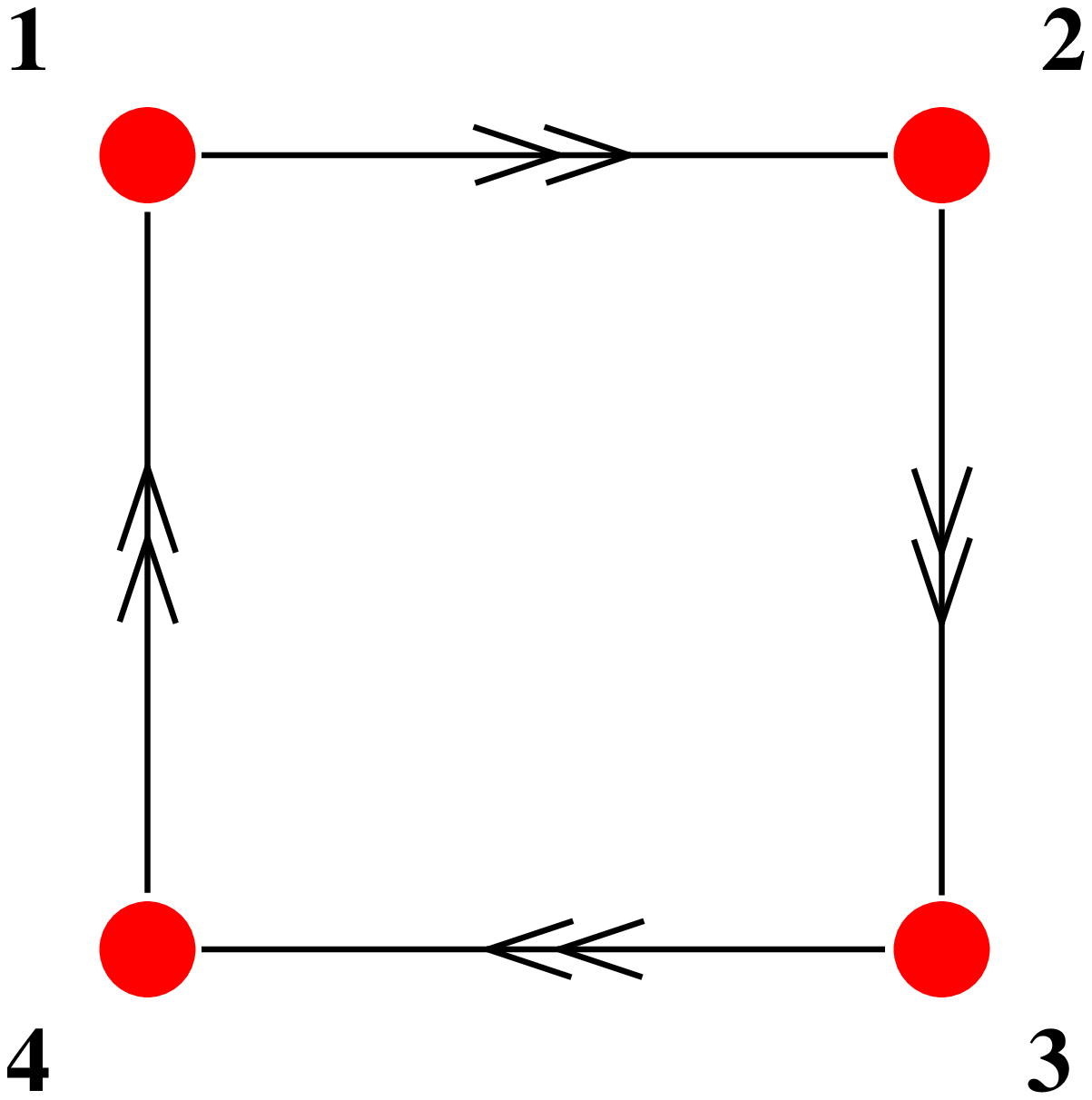}} \ea & \ba{c} {\cal I}_{II}=\left(\begin{array}{cccc} 0 & -2 & 0 & 2 \\ 2 & 0 & -2 & 0 \\ 0 & 2 & 0 & -2 \\ -2 & 0 & 2 & 0 \end{array} \right) \ea \\

\hline
\end{array}
\eeq

\bigskip

\underline{Cone over $dP_0$}

\medskip

$dP_0$ has one phase with 9 fields.

\beq
\begin{array}{|c|c|c|c|}
\hline
(p,q) \ web & Quiver & Intersection \ matrix & \begin{array}{c} Fractional \ brane \\ charges \end{array} \\
\hline
\ba{c} {\epsfxsize=2.7cm\epsfbox{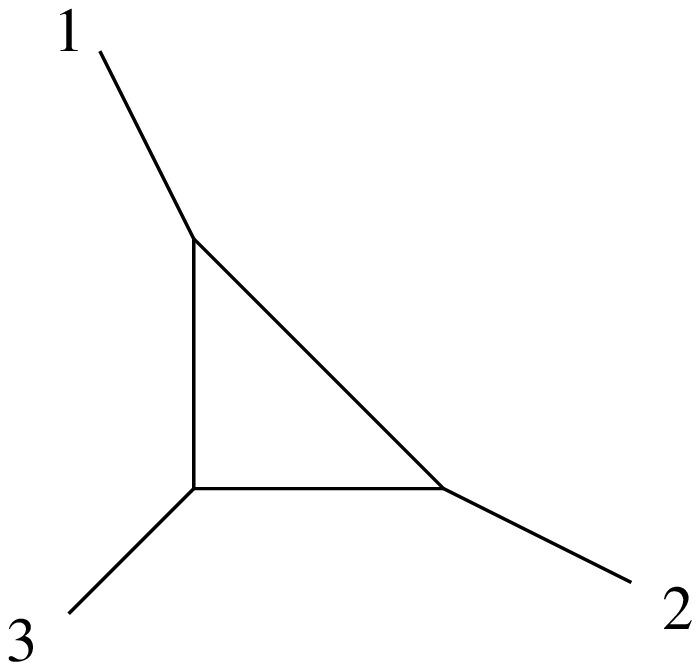}} \ea & \ba{c} {\epsfxsize=3cm\epsfbox{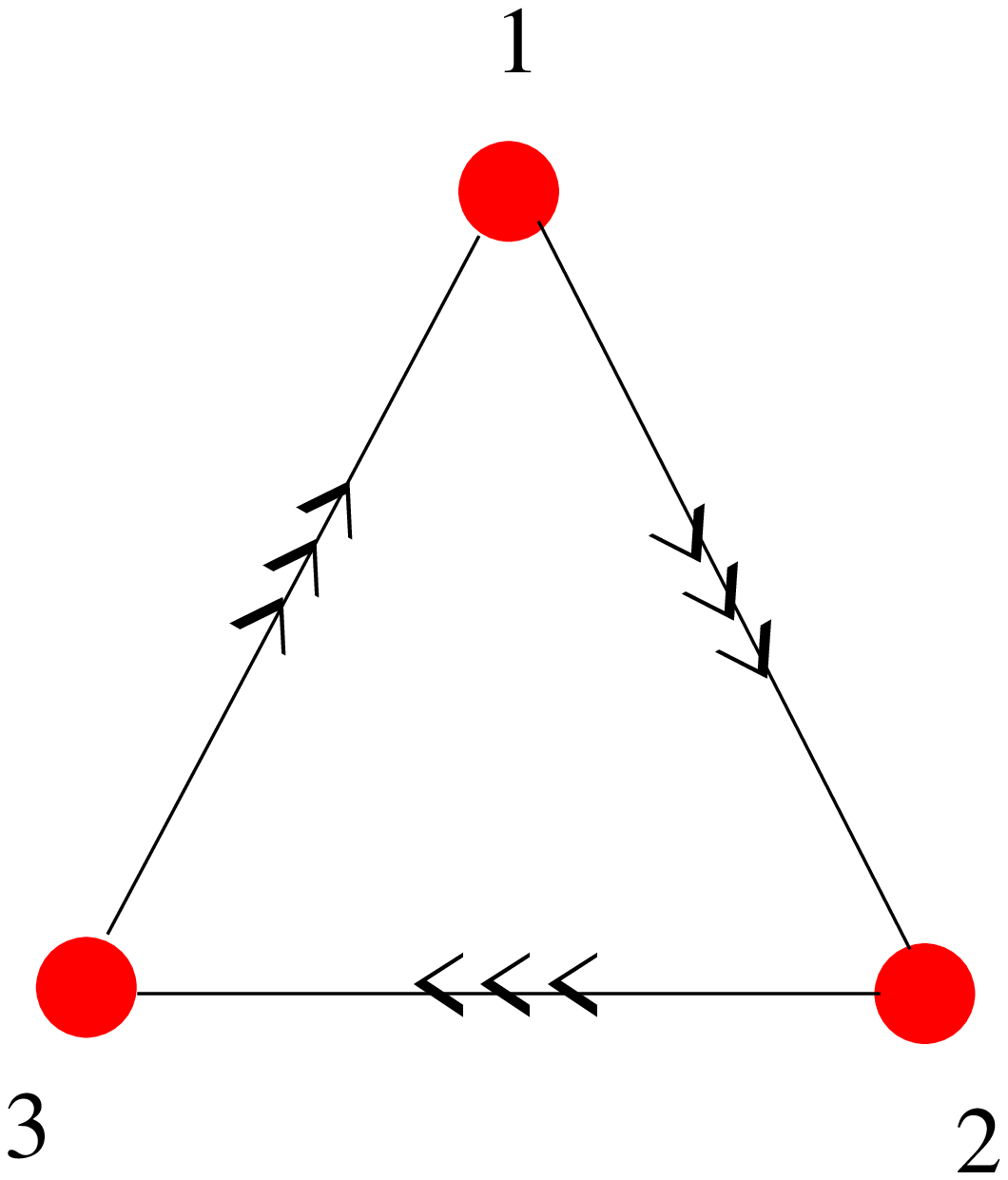}} \ea & \small{ \ba{c} {\cal I}=\left(\begin{array}{ccc} 0 & -3 & 3 \\ 3 & 0 & -3  \\ -3 & 3 & 0 
\end{array} \right) \ea } &

\small{
\begin{array}{l}
ch(F_1)=(2,-\ell,-1/2) \\
ch(F_2)=(-1,\ell,-1/2) \\
ch(F_3)=(-1,0,0)
\end{array}
}

\\
\hline
\end{array}
\eeq

\newpage

\underline{Cone over $dP_1$}

\medskip

$dP_1$ has one phase with 10 fields.

\beq
\begin{array}{|c|c|c|c|}
\hline
(p,q) \ web & Quiver & Intersection \ matrix & \begin{array}{c} Fractional \ brane \\ charges \end{array} \\
\hline
\ba{c} {\epsfxsize=2.7cm\epsfbox{web_dP1.eps}} \ea & \ba{c} {\epsfxsize=3.7cm\epsfbox{quiver_dP1.eps}} \ea & \small{ \ba{c} {\cal I}=\left(\begin{array}{cccc} 0 & -2 & -1 & 3 \\ 2 & 0 & -1 & -1 \\ 1 & 1 & 0  & -2 \\ -3 & 1 & 2  & 0  \end{array} \right) \ea }&

\small{
\begin{array}{l}
ch(F_1)=(2,-\ell,-1/2) \\
ch(F_2)=(0,E_1,-1/2) \\
ch(F_3)=(-1,\ell-E_1,0) \\
ch(F_4)=(-1,0,0)
\end{array}
}

\\
\hline
\end{array}
\eeq

\bigskip

\underline{Cone over $dP_2$}

\medskip

$dP_2$ has two phases, with 13 and 11 fields.

\beq
\begin{array}{|c|c|c|c|}
\hline
(p,q) \ web & Quiver & Intersection \ matrix & \begin{array}{c} Fractional \ brane \\ charges \end{array} \\
\hline
\ba{c} {\epsfxsize=2.7cm\epsfbox{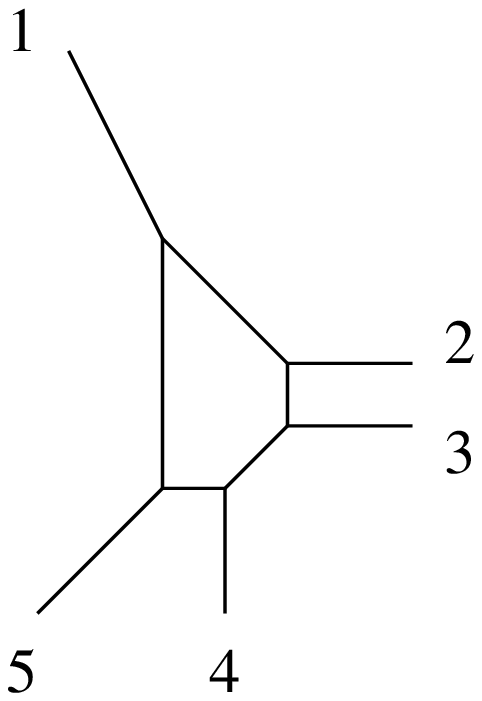}} \ea & \ba{c} {\epsfxsize=3.7cm\epsfbox{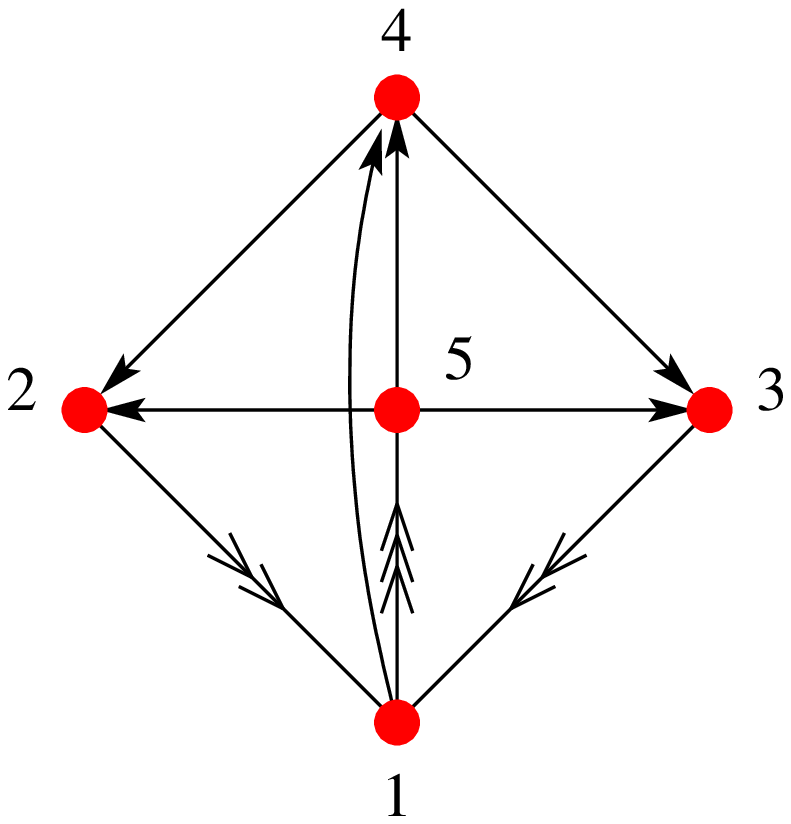}} \ea & \small{ \ba{c} {\cal I}_{I}=\left(\begin{array}{ccccc} 0 & -2 & -2 & 1 & 3 \\ 2 & 0 & 0 & -1 & -1 \\ 2 & 0 & 0 & -1 & -1 \\ -1 & 1 & 1 & 0 & -1 \\ -3 & 1 & 1 & 1 & 0 \end{array} \right) \ea }&

\small{
\begin{array}{l}
ch(F_1)=(2,-\ell,-1/2) \\
ch(F_2)=(0,E_1,-1/2) \\
ch(F_3)=(0,E_2,-1/2) \\
ch(F_4)=(-1,\ell-E_1-E_2,1/2) \\
ch(F_5)=(-1,0,0)
\end{array}
}

\\
\hline
\ba{c} {\epsfxsize=2.7cm\epsfbox{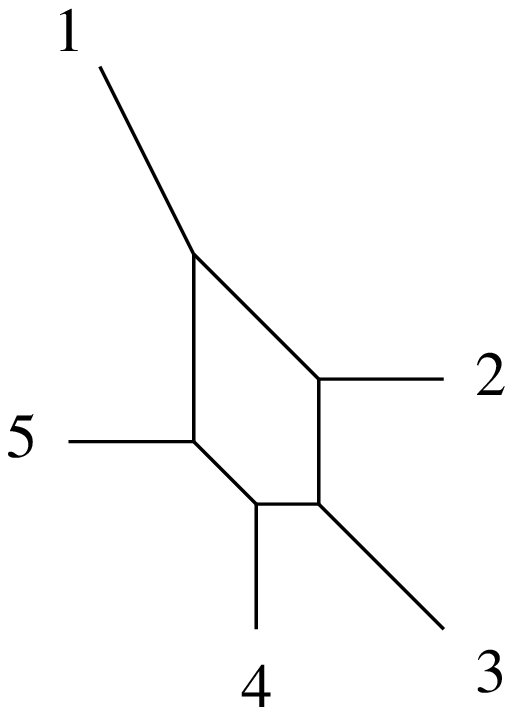}} \ea & \ba{c} {\epsfxsize=3.7cm\epsfbox{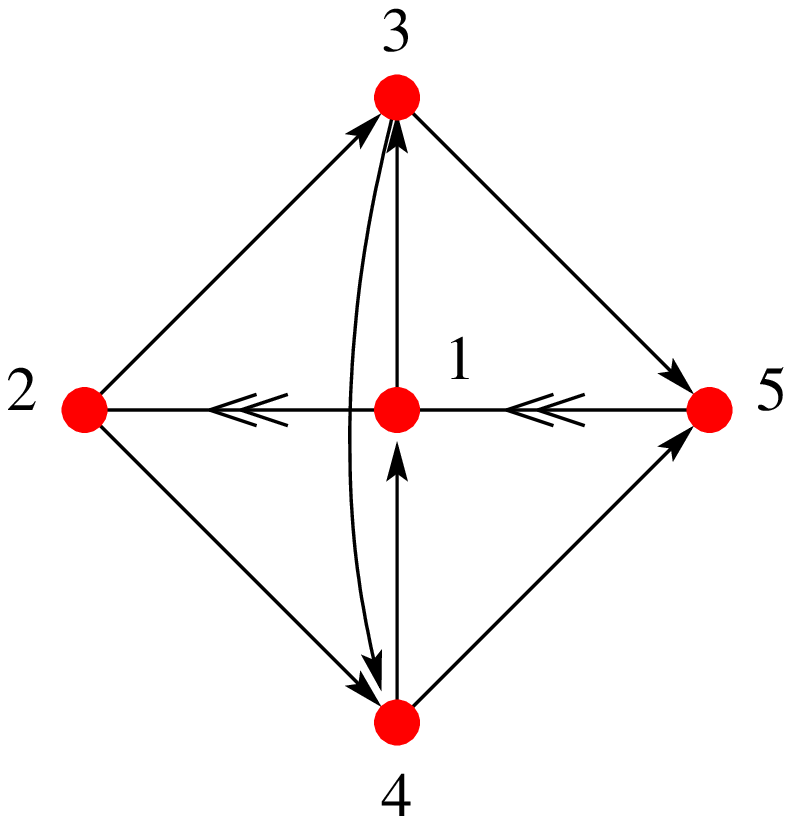}} \ea & \small{ \ba{c} {\cal I}_{II}=\left(\begin{array}{ccccc} 0 & -2 & -1 & 1 & 2 \\ 2 & 0 & -1 & -1 & 0 \\ 1 & 1 & 0 & -1 & -1 \\ -1 & 1 & 1 & 0 & -1 \\ -2 & 0 & 1 & 1 & 0 \end{array} \right) \ea }&

\small{
\begin{array}{l}
ch(F_1)=(2,-\ell,-1/2) \\
ch(F_2)=(0,E_1,-1/2) \\
ch(F_3)=(-1,\ell-E_1,0) \\
ch(F_4)=(-1,E_2,1/2) \\
ch(F_5)=(0,-E_2,-1/2)
\end{array}
}

\\
\hline
\end{array}
\eeq

\newpage

\underline{Cone over $dP_3$}

\medskip

$dP_3$ has four phases, with 12, 14, 14 and 18 fields.

\beq
\begin{array}{|c|c|c|c|}
\hline
(p,q) \ web & Quiver & Intersection \ matrix  & \begin{array}{c} Fractional \ brane \\ charges \end{array} \\
\hline
\ba{c} {\epsfxsize=2.7cm\epsfbox{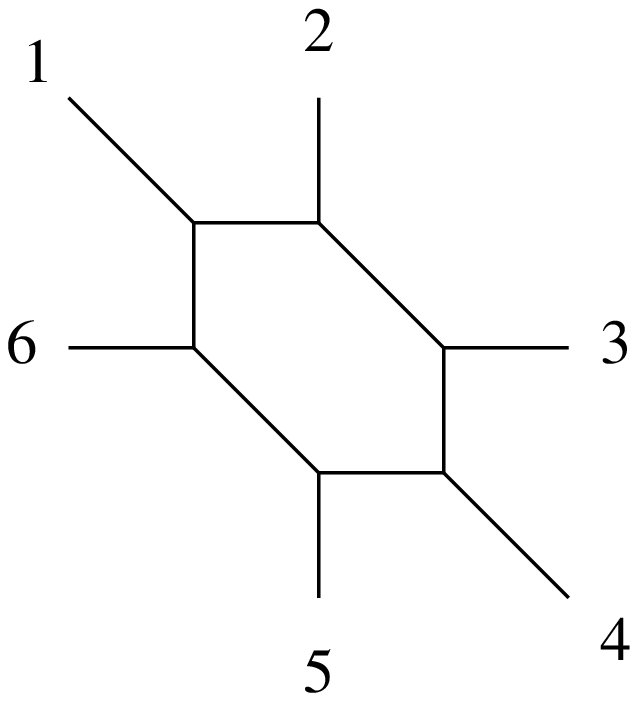}} \ea & \ba{c} {\epsfxsize=3cm\epsfbox{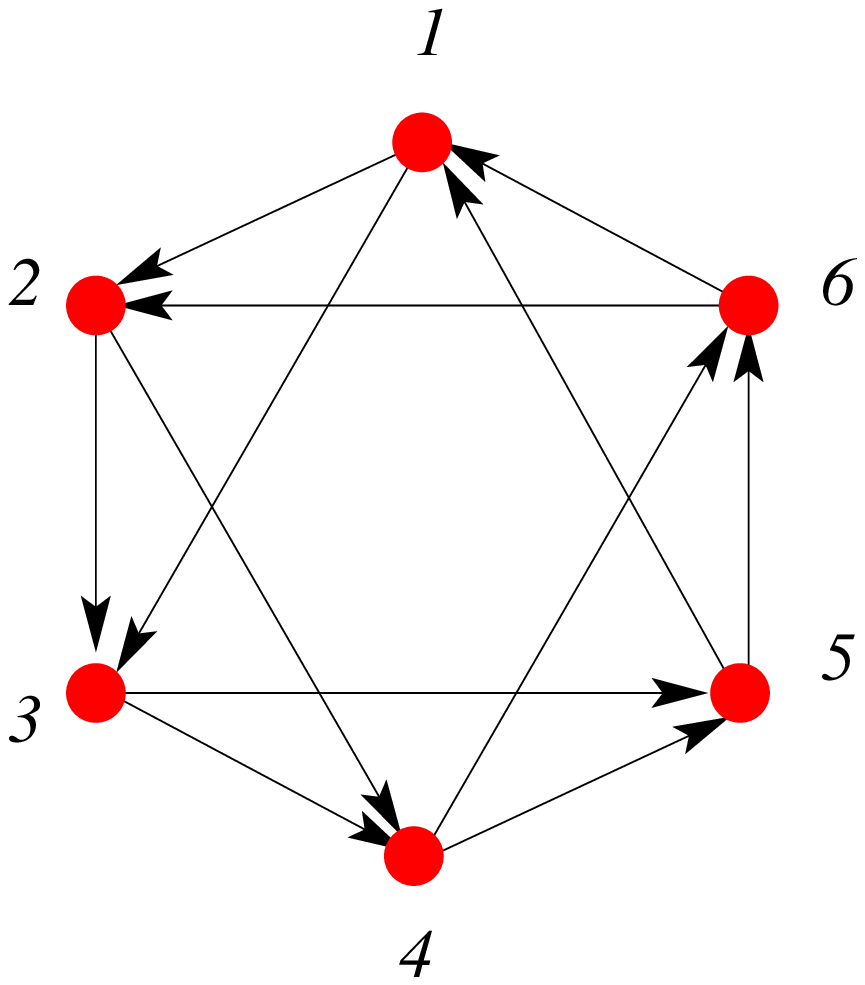}} \ea & \small{ \ba{c} {\cal I}_I=\left(  \begin{array}{cccccc} 0 & -1 & -1 & 0 & 1 & 1 \\ 1& 0 & -1 & -1 & 0 & 1 \\ 1 & 1 & 0 & -1 & -1 & 0 \\ 0 & 1 & 1 & 0 & -1 & -1 \\  -1 & 0 & 1 & 1 & 0 & -1 \\ -1 & -1 & 0 & 1 & 1 & 0  \end{array} \right) \ea }&

\small{
\begin{array}{l}
ch(F_1)=(1,-\ell+E_3,0) \\
ch(F_2)=(1,-E_3,-1/2) \\
ch(F_3)=(0,E_1,-1/2) \\
ch(F_4)=(-1,\ell-E_1,0) \\
ch(F_5)=(-1,E_2,1/2) \\
ch(F_6)=(0,-E_2,-1/2)
\end{array}
}

\\
\hline
\ba{c} {\epsfxsize=2.7cm\epsfbox{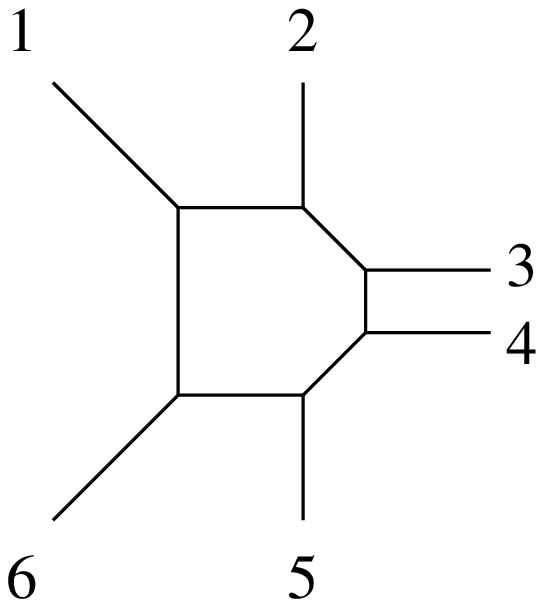}} \ea & \ba{c} {\epsfxsize=3cm\epsfbox{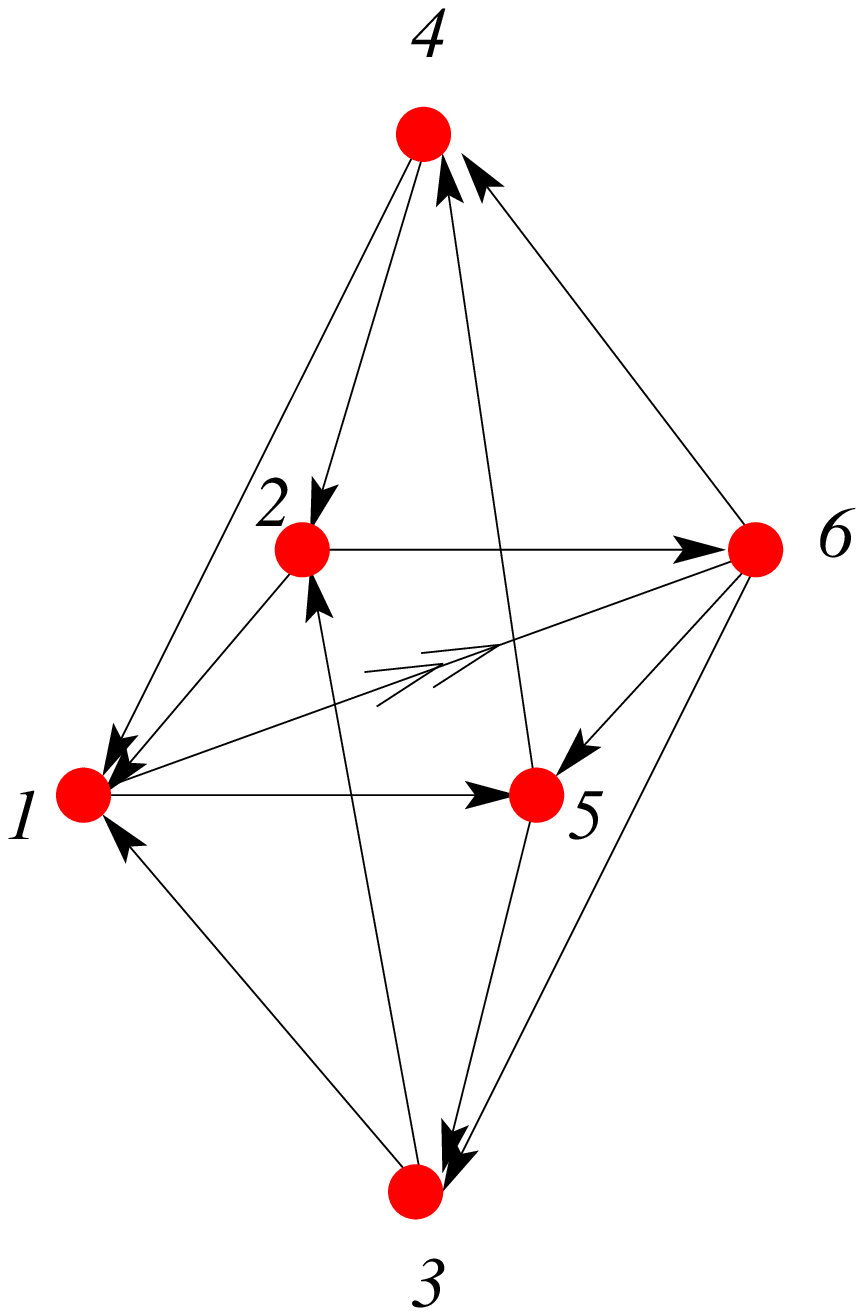}} \ea & \small{ \ba{c} {\cal I}_{II}=\left(  \begin{array}{cccccc}0 & -1 & -1 & -1 & 1 & 2 \\ 1 & 0 & -1 & -1 & 0 & 1 \\ 1 & 1 & 0 & 0 & -1 & -1 \\ 1 & 1 & 0 & 0 & -1 & -1 \\ -1 & 0 & 1 & 1 & 0 & -1 \\ -2 & -1 & 1 & 1 & 1 & 0 \end{array} \right) \ea }&

\small{
\begin{array}{l}
ch(F_1)=(1,-\ell+E_3,0) \\
ch(F_2)=(1,-E_3,-1/2) \\
ch(F_3)=(0,E_1,-1/2) \\
ch(F_4)=(0,E_2,-1/2) \\
ch(F_5)=(-1,\ell-E_1-E_2,1/2) \\
ch(F_6)=(-1,0,0)
\end{array}
}

\\
\hline
\ba{c} {\epsfxsize=2.7cm\epsfbox{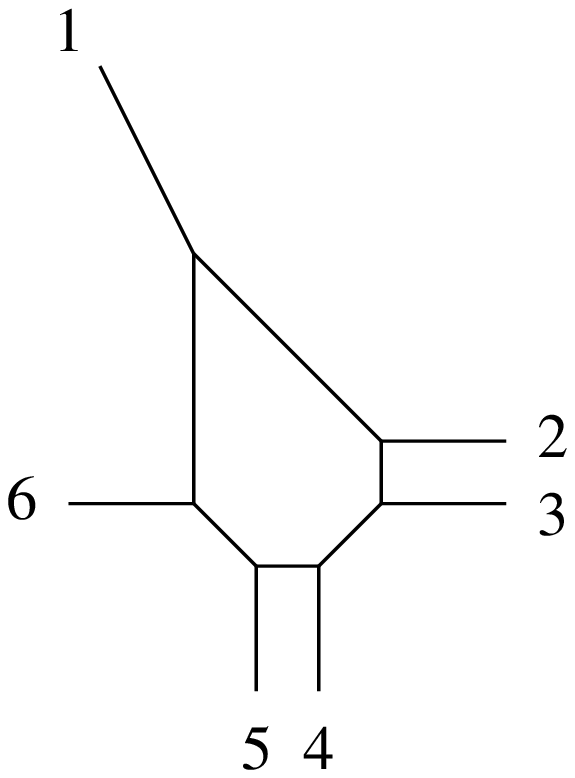}} \ea & \ba{c} {\epsfxsize=3cm\epsfbox{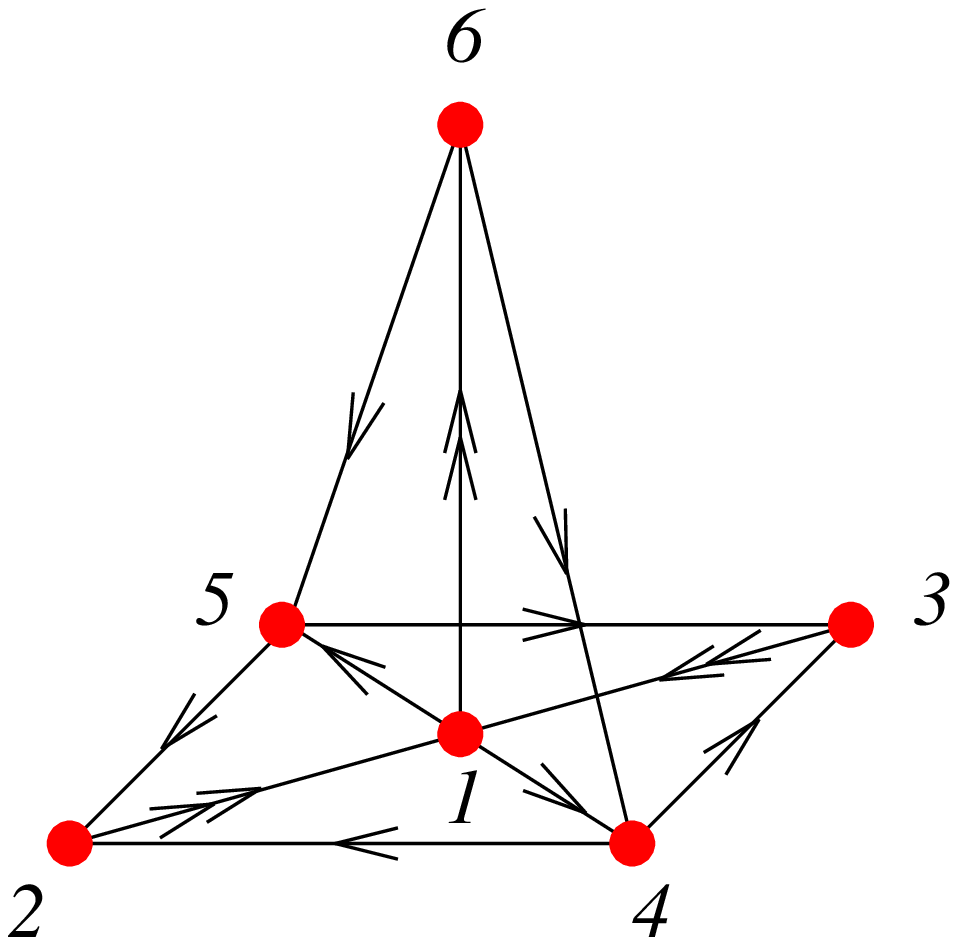}} \ea &  \small{ \ba{c} {\cal I}_{III}=\left(  \begin{array}{cccccc}0 & -2 & -2 & 1 & 1 & 2 \\ 2 & 0 & 0 & -1 & -1 & 0 \\ 2 & 0 & 0 & -1 & -1 & 0 \\ -1 & 1 & 1 & 0 & 0 & -1 \\ -1 & 1 & 1 & 0 & 0 & -1 \\ -2 & 0 & 0 & 1 & 1 & 0 \end{array} \right) \ea }&

\small{
\begin{array}{l}
ch(F_1)=(2,-\ell,-1/2) \\
ch(F_2)=(0,E_1,-1/2) \\
ch(F_3)=(0,E_3,-1/2) \\
ch(F_4)=(-1,\ell-E_1-E_3,1/2) \\
ch(F_5)=(-1,E_2,1/2) \\
ch(F_6)=(0,-E_2,-1/2)
\end{array}
}

\\
\hline
\ba{c} {\epsfxsize=2.7cm\epsfbox{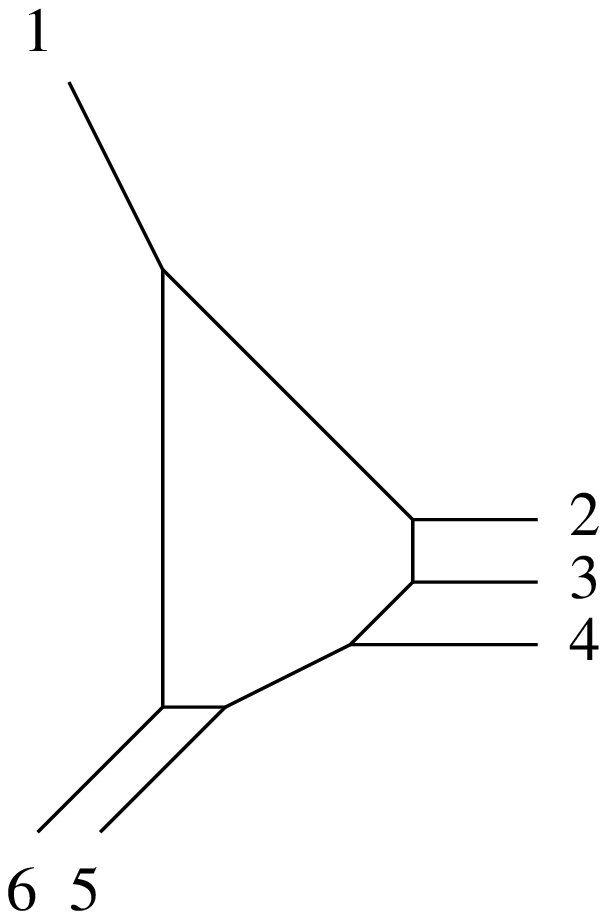}} \ea & \ba{c} {\epsfxsize=3cm\epsfbox{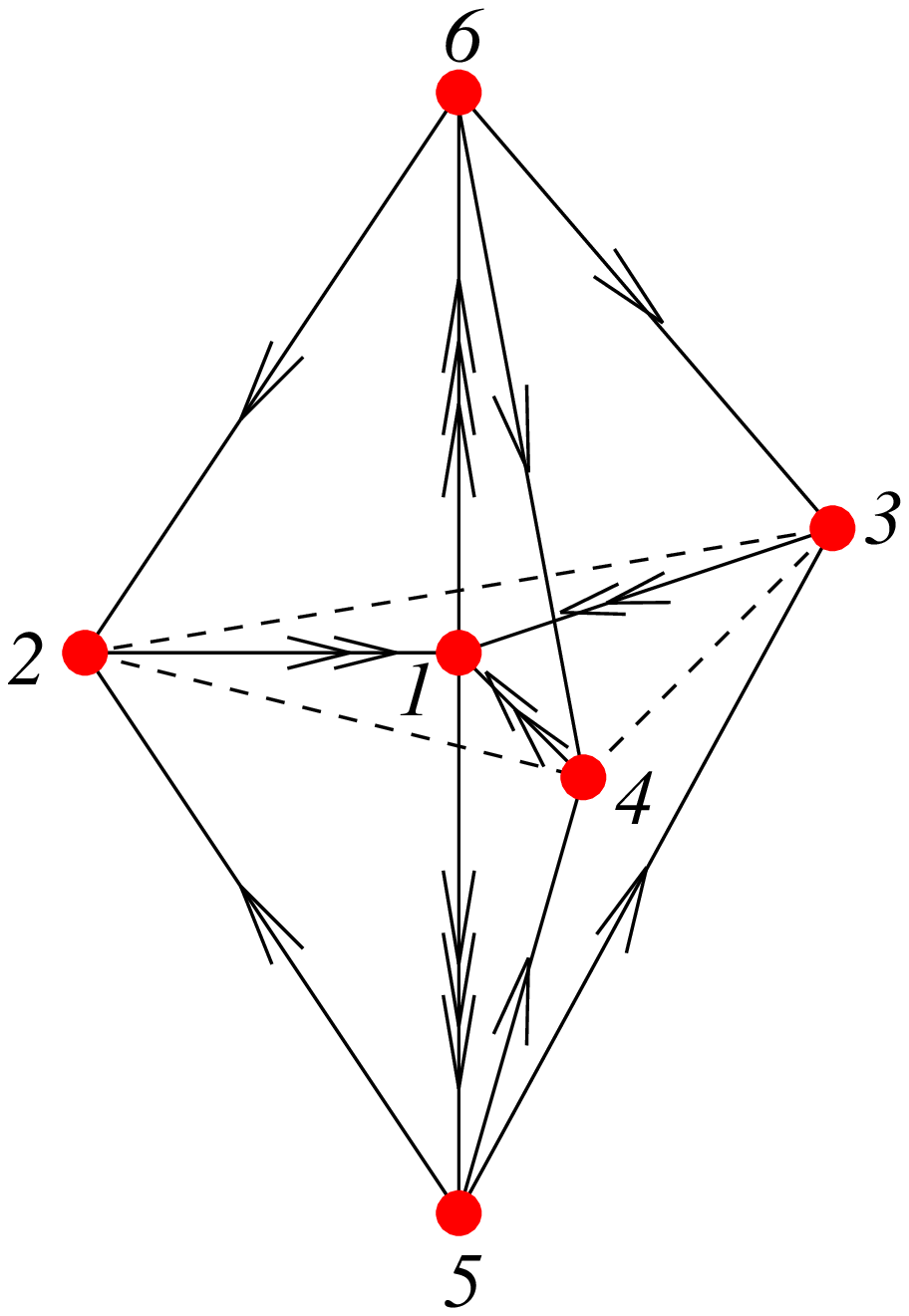}} \ea &  \small{ \ba{c} {\cal I}_{IV}=\left(  \begin{array}{cccccc}0 & -2 & -2 & -2 & 3 & 3 \\ 2 & 0 & 0 & 0 & -1 & -1 \\ 2 & 0 & 0 & 0 & -1 & -1\\ 2 & 0 & 0 & 0 & -1 & -1 \\ -3 & 1 & 1 & 1 & 0 & 0 \\ -3 & 1 & 1 & 1 & 0 & 0 \end{array} \right) \ea }&

\small{
\begin{array}{l}
ch(F_1)=(2,-\ell,-1/2) \\
ch(F_2)=(0,E_1,-1/2) \\
ch(F_3)=(0,E_2,-1/2) \\
ch(F_4)=(0,E_3,-1/2) \\
ch(F_5)=(-1,\ell-E_1-E_2-E_3,1) \\
ch(F_6)=(-1,0,0)
\end{array}
}

\\
\hline
\end{array}
\eeq


\bibliographystyle{JHEP}

\end{document}